\documentclass[superscriptaddress, twocolumn, amsmath, amssymb, aps,
pra, letterdraft, notitlepage]{revtex4-1} 

\usepackage{float}
\makeatletter
\let\newfloat\newfloat@ltx
\makeatother
\usepackage{algorithm,algcompatible,amsmath}
\algnewcommand\INPUT{\item[\textbf{Input:}]}%
\algnewcommand\OUTPUT{\item[\textbf{Output:}]}%

\usepackage{booktabs}
\usepackage[english]{babel}

\usepackage{graphicx,graphics,epsfig,subfigure,times,bm,bbm,amssymb,amsmath,amsfonts,amsthm,mathrsfs,MnSymbol}
\usepackage[matrix,frame,arrow]{xypic}
\usepackage[normalem]{ulem}
\usepackage{slashed}
\usepackage{color}
\usepackage[usenames,dvipsnames,svgnames,table]{xcolor}
\usepackage[english]{babel}
\definecolor{darkblue}{rgb}{0.0,0.0,0.3}
\usepackage[colorlinks=true,
            linkcolor=red,
            urlcolor= darkblue,
            citecolor=blue]{hyperref}

\newcommand{\bes} {\begin{subequations}}
\newcommand{\ees} {\end{subequations}}
\newcommand{\bea} {\begin{eqnarray}}
\newcommand{\eea} {\end{eqnarray}}

\def\>{\rangle}
\def\<{\langle}

\newcommand{\ket}[1]{|#1\rangle}

\begin{document}

\title{A non-Hermitian Ground State Searching Algorithm Enhanced by Variational Toolbox}

\author{Yu-Qin Chen}
\affiliation{Tencent Quantum Laboratory, Tencent, Shenzhen, Guangdong, China, 518057}
\author{Shi-Xin Zhang}
\affiliation{Tencent Quantum Laboratory, Tencent, Shenzhen, Guangdong, China, 518057}
\author{Chang-Yu Hsieh}
\email{kimhsieh2@gmail.com}
\affiliation{Tencent Quantum Laboratory, Tencent, Shenzhen, Guangdong, China, 518057}
\author{Shengyu Zhang}
\email{shengyzhang@tencent.com}
\affiliation{Tencent Quantum Laboratory, Tencent, Shenzhen, Guangdong, China, 518057}

\begin{abstract}
Ground-state preparation for a given Hamiltonian is a common quantum-computing task of great importance and has relevant applications in quantum chemistry, computational material modeling, and combinatorial optimization. We consider an approach to simulate dissipative non-Hermitian Hamiltonian quantum dynamics using Hamiltonian simulation techniques to efficiently recover the ground state of a target Hamiltonian. The proposed method facilitates the energy transfer by repeatedly projecting ancilla qubits to the desired state, rendering the effective non-Hermitian Hamiltonian evolution on the system qubits. 
To make the method more resource friendly in the noisy intermediate-scale quantum (NISQ) and early fault-tolerant era, we combine the non-Hermitian projection algorithm with multiple variational gadgets, including variational module enhancement and variational state recording, to reduce the required circuit depth and avoid the exponentially vanishing success probability for post-selections.  We compare our method, the non-Hermitian-variational algorithm, with a pure variational method - QAOA for solving the 3-SAT problem and preparing the ground state for the transverse-field Ising model.  As demonstrated by numerical evidence, the  non-Hermitian-variational algorithm outperforms QAOA in convergence speed with improved quantum resource efficiency.


\end{abstract}

\maketitle
\section{Introduction}
Quantum simulation is one of the most anticipated applications of quantum computation. The goal of quantum simulation is to quantitatively determine the dynamics and physical properties of many-body quantum systems, which is expected to manifest exponential speedup over classical simulation schemes~\cite{feynman2018simulating,lanyon2010towards,wecker2015solving}. 
However, the quantum computation in the near future is performed on the so-called noisy intermediate-scale quantum (NISQ) hardware~\cite{preskill2018quantum}, characterized by shallow circuit depths, limited error-mitigating capability and no large-scale error corrections. Suitable simulation algorithms able to efficiently exploit the presently scarce and error-prone quantum computing resources to simulate non-trivial quantum systems are urgently called for developments~\cite{moll2018quantum,cross2019validating}.


In the NISQ era, the variational method is one of the most popular approaches to prepare the ground state of a given problem Hamiltonian $H_S$. A prototypical example is variational quantum eigensolver (VQE)~\cite{peruzzo2014variational,wang2015quantum,o2016scalable,shen2017quantum,mcclean2016theory,paesani2017experimental}, which relies on a quantum-classical hybrid optimization loop to iteratively tune parameterized gates in order to realize a variational ansatz that minimizes the energy. VQE has become the method of choice for exploring a wide range of tasks from quantum chemistry~\cite{moll2018quantum,kandala2017hardware,cao2019quantum}, condensed matter~\cite{bravo2020scaling,mcardle2020quantum,verstraete2009quantum,schmoll2017kitaev,hebenstreit2017compressed, zhang2022variational, zhang2021variational, liu2021probing,chen2021variational}, and combinatorial optimization~\cite{garcia2018addressing}. Conceptually, VQE is attractive as a shallow parameterized quantum circuit can still potentially encode a highly non-trivial quantum state hard to simulate classically. However, practical challenges of optimizing a vast number of circuit parameters have proven to be formidable~\cite{Rudolph2021}. Many optimization algorithms, such as gradient descent, are liable to becoming trapped in local minima~\cite{mcclean2018barren,diez2021quantum} of a high-dimensional energy landscape, and general problem agnostic circuit ansatz can be plagued with the barren plateaus~\cite{tilly2021variational,mcclean2018barren,wang2021noise}. 

Quantum approximate optimization algorithm (QAOA)~\cite{farhi2014quantum,lloyd2018quantum,kadowaki1998quantum,morales2020universality,morales2020universality,harrigan2021quantum} is another extensively studied variational quantum algorithm. 
 QAOA utilizes a problem-dependent ansatz architecture inspired by the adiabatic quantum computation~\cite{farhi2000quantum} as it constructs the circuit by alternating parameterized unitary circuits generated by either a mixing Hamiltonian $H_B$ or the problem Hamiltonian $H_S$. The ground state of $H_S$ could be reached via properly tuned parameters when the ansatz circuit is deep enough.  Although various techniques to improve QAOA have been proposed~\cite{yang2017optimizing, venturelli2018compiling, oddi2018greedy, zhang2022variational}, due to its variational nature, only a few insights exist in terms of the performance and the scaling behavior for QAOA~\cite{zhou2020quantum,hastings2019classical,streif2019comparison}. The task of identifying optimal parameters of the QAOA circuit becomes more challenging as the circuit depth increases~\cite{zhou2020quantum}.

Imaginary time evolution (ITE) approaches~\cite{lehtovaara2007solution,kraus2010generalized,mcclean2015compact} utilize evolution functions as an energy filter that gradually filters out high-energy part of the initial state and eventually recovers the ground state. Unlike variational approaches, ITE approaches are theoretically reliable and efficient for the ground-state simulation, as long as the overlap of the initial state and the target state is not exponentially small. Different from VQE and QAOA, ITE-based simulation algorithms do not rely on variational ansatz and thus are free from challenges of classical optimizations. However, imaginary-time propagator cannot be straightforwardly decomposed into a unitary gates sequence, namely, ITE approach is generally impossible to directly realized on a quantum computer using polynomial resources~\cite{abrams1997simulation}, though many attempts are proposed~\cite{mcardle2019variational,motta2020determining,yeter2021benchmarking}.  Besides conventional imaginary time evolution, quantum simulation algorithms based on other kinds of energy filters also attract more academic attention recently~\cite{trefethen1997numerical,noble2013generalized,banuls2020entanglement,kitaev1995quantum,poulin2009preparing,abrams1999quantum,ge2019faster,lu2021algorithms,amaro2021filtering}. Similar as the ITE approach, all energy-filtering algorithms are theoretically guaranteed to obtain the ground state by gradually filtering out high-energy components though consume much more quantum computational resources than variational approaches. 
Cosine filter~\cite{ge2019faster,lu2021algorithms,amaro2021filtering} is one of the energy-filtering methods believed to significantly outperform the well-established phase estimation scheme~\cite{kitaev1995quantum}. 
Its runtime scales polynomially better with the spectral gap and the overlap of the trial state with the ground state in the allowed error to the real ground state~\cite{ge2019faster}.

In this work, we propose a family of non-Hermitian-dynamics inspired algorithms for the ground-state simulation. We analytically prove that the core idea of the non-Hermitian Hamiltonian evolution algorithm is to provide a straightforward protocol to implement the aforementioned cosine-filter evolution. We then numerically demonstrate the effectiveness of our new algorithm for preparing the ground state of quantum many-body problems as well as classical combinatorial optimization problems.

A major goal of this work is to build upon a theoretically solid simulation methodology and propose techniques utilizing variational gadgets that could reduce the required quantum resources for experiments in the NISQ and early fault-tolerant era. One main technique proposed is to implement a variational-module-assisted non-Hermitian Hamiltonian dynamics on the quantum devices that accelerate the convergence of the dissipative/energy filtering process towards the ground state. This hybridization of traditional Hamiltonian simulation techniques (such as the Trotterization of dynamical propagator $e^{-iHt}$) and the parameterized quantum circuits (with variationally tunable parameters) is of high novelty and importance. On the one hand, it can greatly reduce the circuit depth and post-selection fail trials for Hamiltonian simulation algorithms. On the other hand, such hybrid pipeline can assuage the challenges of optimizing a tremendous number of gate parameters if the simulation was done solely within the variational framework. In this study, we investigate several hybridized circuit structures with non-Hermitian Hamiltonian evolution and variational modules and identify that the alternating layout for non-Hermitian propagator and variational blocks is suitable. We also show that with a very simple variational circuit structure (as simple as one layer of single-qubit rotation gates), the original non-Hermitian propagator can be greatly accelerated as confirmed by experiments. Furthermore, we find that the non-Hermitian-variational hybrid algorithm outperforms the standard variational algorithm QAOA and its variants in several standard tasks.

The second main technique that we incorporate variational idea into the non-Hermitian Hamiltonian simulation algorithm is the state-recording procedure. The non-Hermitian Hamiltonian evolution simulation faces a vital challenge: the energy dissipation for non-Hermitian propagation is implemented by repeatedly post-selecting the measurement results of ancilla qubits. In general, the required computational resources for all post-selections to succeed increase exponentially with the number of post-selections. Via the first technique where two types of circuits are merged together, the algorithm not only accelerates the convergence of the energy-filtering process but also limits the number of ancilla qubits to be post-selected or processed for the same accuracy. Nevertheless, the challenge remains that the algorithm's runtime scales exponentially with the number of ancilla qubits. To further alleviate this critical burden, state recording technique is utilized which exchanges the expenses of quantum computational resources with classical optimization ones. Inspired by previous works~\cite{lin2021real,barison2021efficient}, we break the full evolution time of the non-Hermitian Hamiltonian and record intermediate approximate quantum states with relatively shallow parameterized quantum circuit determined by a partial quantum tomography.  Different from the real-time dynamical simulation in Ref.~\cite{lin2021real,barison2021efficient}, the number of measurements can be drastically minimized in our state recording procedure. Details on two possible partial tomography strategies we proposed (suitable for different types of problems) will be elucidated below.

This paper is organized as follows. We
firstly present the framework of non-Hermitian Hamiltonian simulation algorithm and theoretically show that it effectively realizes the cosine energy filter and can reach the ground state in Sec II.  We then show the numerical results on quantum many-body simulation tasks as well as combinatorial optimization tasks in Sec III.
Next, we discuss a hybrid version of this non-Hermitian simulation algorithm, which introduces variational modules to accelerate the convergence toward the desired ground state in Sec IV. Then we present the method of breaking the full non-Hermitian Hamiltonian evolution into time steps by state recording in Sec V. This state-recording
scheme is introduced to save times of post-selection since the success probability of implementing the evolution decreases exponentially with the number of post-selections and thus time steps. 
Variational state recording consumes extra time and resources to perform. Therefore, we further propose customized strategy of reduced state recording methods for many-body quantum systems and classical combinatorial optimization problem simulation in Sec V to further minimize the required number of measurement shots in real implementations. Finally, in Sec VI,  we discuss the further implication of our algorithms with hybrid modules from fault-tolerant and NISQ paradigms.


\section{non-Hermitian Hamiltonian simulation algorithm}
\begin{figure}[htp]
\centering
\includegraphics[width=0.48\textwidth]{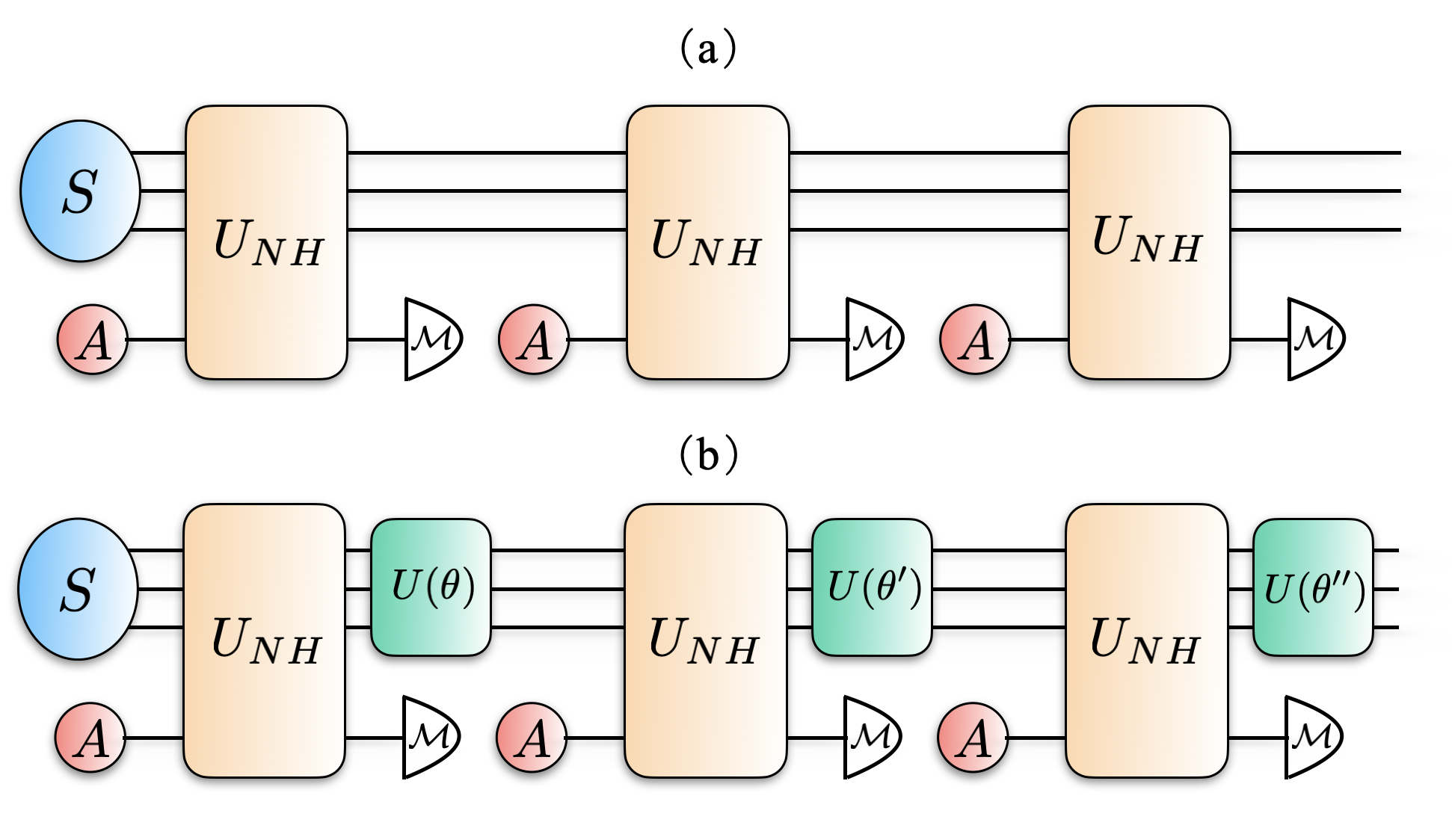}
\caption{(a) The quantum circuit realization of non-Hermitian algorithm. $S$ (blue) is the problem system prepared in state $\left|\psi_0\right>$. $A$ (red) is the ancilla qubit prepared in state $\left|0\right>$.  Non-Hermitian block $U_{NH}$ (yellow) is repeatedly applied where each Hamiltonian simulation block is followed by a post-selection on ancilla qubit to state $\vert 0\rangle$. (b) The quantum circuit realization of non-Hermitian-Variational hybrid algorithm. $U(\theta)$ (green) is the variational block that can be optimized with tuning parameters $\theta$.     }
\label{fig:circuit}
\end{figure}

We propose a non-Hermitian Hamiltonian evolution algorithm that is effectively equivalent to the cosine filter approach~\cite{ge2019faster,lu2021algorithms,amaro2021filtering} for simulating the ground state. The core of the
algorithm is simply the Hamiltonian simulation, which is a theoretically well-established algorithm primitive that manifests quantum advantage~\cite{wiesner1996simulations,lloyd1996universal,abrams1997simulation,zalka1998simulating,kassal2008polynomial}. The non-Hermitian Hamiltonian simulation can be realized on a quantum computer as sketched in Fig.\ref{fig:circuit}. 

We start from an initial state $\left|\psi_0\right>$ in $n$-qubit system, coupled to an ancilla qubit $\left|0\right>$. Then we construct a real-time evolution block denoted by $U_{NH}$, 
\begin{eqnarray}
U_{NH}=e^{-iHs\otimes Y_A dt},
\end{eqnarray}
where $H_S$ is the problem Hamiltonian, $Y_A=\left(\begin{array}{cc}0 & -i \\ i & 0\end{array}\right)$ is the Pauli-Y matrix on the ancilla qubit, $dt$ is the time for the Hamiltonian propagator.  
We repeatedly apply the non-Hermitian block $U_{NH}$ $M$ times with a new ancilla qubit each time step, then the system under Hamiltonian evolved with time $Mdt$ is
\begin{equation}
\begin{array}{l}\left|\phi_{M}\right\rangle=U_{NH}^{M}\left|\psi_{0}\right\rangle|00 \ldots 0\rangle \\ =\cos ^{M}\left(H_{S} dt\right)\left|\psi_{0}\right\rangle|00 \ldots 0\rangle  \\+\cos ^{\mathrm{M}-1}\left(H_{S} dt\right) \cdot \sin \left(H_{S} dt\right)\left|\psi_{0}\right\rangle|00 \ldots 1\rangle\\+~~\cdots \\
+\cos\left(H_{S} dt\right) \cdot \sin^{M-1} \left(H_{S} dt\right)\left|\psi_{0}\right\rangle|01 \ldots 1\rangle\\+\sin ^{M}\left(H_{S} dt\right)\left|\psi_{0}\right\rangle|11 \ldots 1\rangle.
\end{array}
\end{equation}
If we project the ancilla qubits to quantum state $\left|00...0\right>$, we obtain the normalized final state,
\begin{equation}
\left|\psi_M\right>=\frac{\cos ^{M}\left(H_{S} dt\right)\left|\psi_{0}\right\rangle}{\| cos^M(H_S dt)\left|\psi_0\right> \|}.
\end{equation}
This corresponds to the cosine-filtering operation introduced in Ref.~\cite{ge2019faster}. Note 
if we post-select the ancilla qubit before applying the next non-Hermitian evolution block, then we can recycle the ancilla qubit to effectively realize an $M$-step time evolution with only one physical ancilla qubit. To show why the cosine filter acts as an energy-filtering operation toward the ground state, we Talyor expand the cosine filter and keep the first nontrivial order of small time $dt$. We then find that its short-time behavior mimics an imaginary-time evolution that targets the desired ground state:
\begin{equation}
\cos ^{M}\left(H_{s} dt\right) \simeq e^{-\left(H_{s} dt\right)^{2} / 2 M}.
\end{equation}

In practice, the Hamiltonian takes the following form:  $H_S=\sum_{n=1}^{N_h}{\alpha_n h_n}$, where $h_n$ are the Pauli strings and $\alpha_n$ are the corresponding coefficients. The minimal and maximum eigenvalues for $H_S$ are denoted as $E_{min}$ and $E_{max}$.  We normalize the Hamiltonian such that the spectrum $|E_{min}|, |E_{max}|\leq N$, so that the spectrum of $H_S$ is strictly confined to the interval $[-N,N]$.  We make a global energy shift to ensure $H_S'=H_S+N/2$ a positive-definite operator, thus the minimal and maximum eigenvalue becomes $E'_{min}>0$,$E'_{max}<N$. 
The evolution time interval required $dt^{\star}$ can be chosen such that $E'_{max}\cdot dt^{\star}\leq \pi/2$ and the effect of the cosine filter is restricted to the region $\left[0,1\right]$. Repeated applications of such a fine-tuned cosine filter are then guaranteed to act as an energy filter that eventually distills the quantum state to the ground state of $H^\prime_S$. More theoretical details on the cosine filter can be found in Appendix \ref{sec: A}.

By post-selecting all the ancilla qubits, we obtain the target system state $\vert \psi_M\rangle$ and can measure the Hamiltonian expectation as $E_M = \langle \psi_M\vert H_S \vert \psi_M\rangle$. Equivalently, we calculate the expected energy for $H_S$ by classical post-processing with the following formula from state before post-selection, 
\begin{equation}\label{eq:processing}
E_{M}=\frac{\left\langle\phi_{M}\left|H_{S} \mathcal{P}\right| \phi_{M}\right\rangle}{\left\langle\phi_{M}|\mathcal{P}| \phi_{M}\right\rangle},
\end{equation}
where $\mathcal{P}=\frac{1}{2^M}(1+Z_1)(1+Z_2)...(1+Z_M)$ is a collective projection operator. Note that the projection operator only contains Pauli-$Z$ terms, which does not introduce additional measurement terms as we can measure these Pauli-$Z$ for ancilla qubits in parallel with the Pauli strings constituting the system Hamiltonian. Furthermore, based on the two facts 1) $\mathcal{PP=P}$ and 2) $\left[\mathcal{P},H_S\right]=0$, we prove that the post-processing evaluation can be cast in the following form, which implies a combinatorial reduction of Pauli string terms (compared to the definition of $\mathcal{P}$),
\begin{equation}\label{eq:pprocessing}
E_{M}=\frac{\mathcal{C}_M^0\left\langle\phi_{0}\left|H_{S}\right| \phi_{0}\right\rangle+\sum_{i=1}^M \mathcal{C}_M^i \left\langle\phi_{i}\left|H_{S}Z_1Z_2\cdots Z_i\right| \phi_{i}\right\rangle }{\mathcal{C}_M^0\left\langle\phi_{0}\vert \phi_{0}\right\rangle+\sum_{i=1}^M \mathcal{C}_M^i \left\langle\phi_{i}\left|Z_1Z_2\cdots Z_i\right| \phi_{i}\right\rangle}.
\end{equation}
Further details on post-selection and post-processing perspectives can be found in Appendix \ref{sec: A}. According to Eq.~\ref{eq:pprocessing}, we only need to measure the following observables in  step $i$:
\begin{equation}\label{eq:addterms}
\left\langle\phi_{i}\left|H_{S}Z_1Z_2\cdots Z_i\right| \phi_{i}\right\rangle,\ \left\langle\phi_{i}\left|Z_1Z_2\cdots Z_i\right| \phi_{i}\right\rangle. 
\end{equation}
Then we can use the observable measured in earlier timestep along with these additional terms in Eq.~\ref{eq:addterms} to construct the energy $E_M$ in Eq.~\ref{eq:pprocessing}. It is worth noting that although we only need to measure polynomial Pauli string terms to give the estimation on the energy, the required total number of measurement shots to given accuracy can still be exponentially large. This complexity is intrinsically induced by the fact that the denominator of  Eq. \ref{eq:pprocessing} is in general exponentially small.

\section{Numerical results for quantum simulation and combinatorial optimization tasks}

In this work, we consider the one-dimensional transverse-field Ising model (TFIM) ground state problem and 3-SAT problems~\cite{Hogg2003,nidari2005,Kirkpatrick1994,Monasson1999} to benchmark our algorithm. The 1D TFIM Hamiltonian reads
\begin{equation}
H_{TFIM}=J \sum_{\langle i j\rangle} Z_{i} Z_{j}+h_{X} \sum X_{i},
\end{equation}
where $X$ and $Z$ are the Pauli matrices and $\left< i,j \right>$ denotes pairs of interacting neighboring qubits, and $J, h_X$ is the coupling strength and transverse field strength, respectively. More model details of the model used in this work are presented in Appendix.~\ref{sec:TFIMmodels}.

The 3-SAT problem is a paradigmatic example of a non-deterministic polynomial (NP) complete problem  \cite{Hogg2003}. A 3-SAT problem is defined by a logical statement involving \(n\) boolean variables \(b_i\). The logical statement consists of \(m\) clauses \(C_i\) in conjunction: \(C_{1} \wedge C_{2} \wedge \cdots \wedge C_{m}\). Each clause is a disjunction of 3 literals, where a literal is a boolean variable \(b_i\) or its negation \(\neg b_i\). For instance, a clause may read \(\left(b_{j} \vee \neg b_{k} \vee b_{l}\right)\). The task is to first decide whether a given 3-SAT problem is satisfiable; if so, then assign appropriate binary values to satisfy the logical statement.  
We can map a 3-SAT problem to the ground state problem for a Hamiltonian on a set of qubits. Under this mapping, each binary variable $b_i$ is represented as a qubit state. Thus, an \(n\)-variable 3-SAT problem is mapped into a Hilbert space of dimension \(N = 2^n\). Furthermore, each clause of the logical statement is translated to a projector, projecting on the bitstrings that not satisfying each given clause. Hence, a logical statement with $m$ clauses may be translated to the following Hamiltonian with long range interaction,
\begin{equation}
H_{3-SAT}=\sum_{\alpha=1}^{m}\left|b_{j}^{\alpha} b_{k}^{\alpha} b_{l}^{\alpha}\right\rangle\left\langle b_{j}^{\alpha} b_{k}^{\alpha} b_{l}^{\alpha}\right|.
\end{equation}
Since the computational complexity is defined in terms of the worst-case performance, hard instances of 3-SAT have been intensively studied in the past. Following Ref \cite{nidari2005}, we focus on a particular set of 3-SAT instances, each characterized by a unique solution and has a ratio $m/n=3$. We note that this ratio of $3$ is different from the phase-transition point $m/n \approx 4.2$  \cite{Kirkpatrick1994,Monasson1999} that has been extensively explored in studies that characterize the degrees of satisfiability of random 3-SAT problems. The subtle distinction is that the phase-transition point characterizes the notion of ``hardness" (with respect to the $m/n$ ratio) by averaging over 3-SAT instances having variable number of solutions. However, when the focus is to identify the most difficult 3-SAT instances having a unique solution, it has been ``empirically" found that these instances tend to have an $m/n$ ratio lower than the phase-transition point. More model details of the combinatorial optimization model used in this work are presented in Appendix ~\ref{sec:3satmodels}.

\begin{figure}[htp]
\centering
\includegraphics[width=0.45\textwidth]{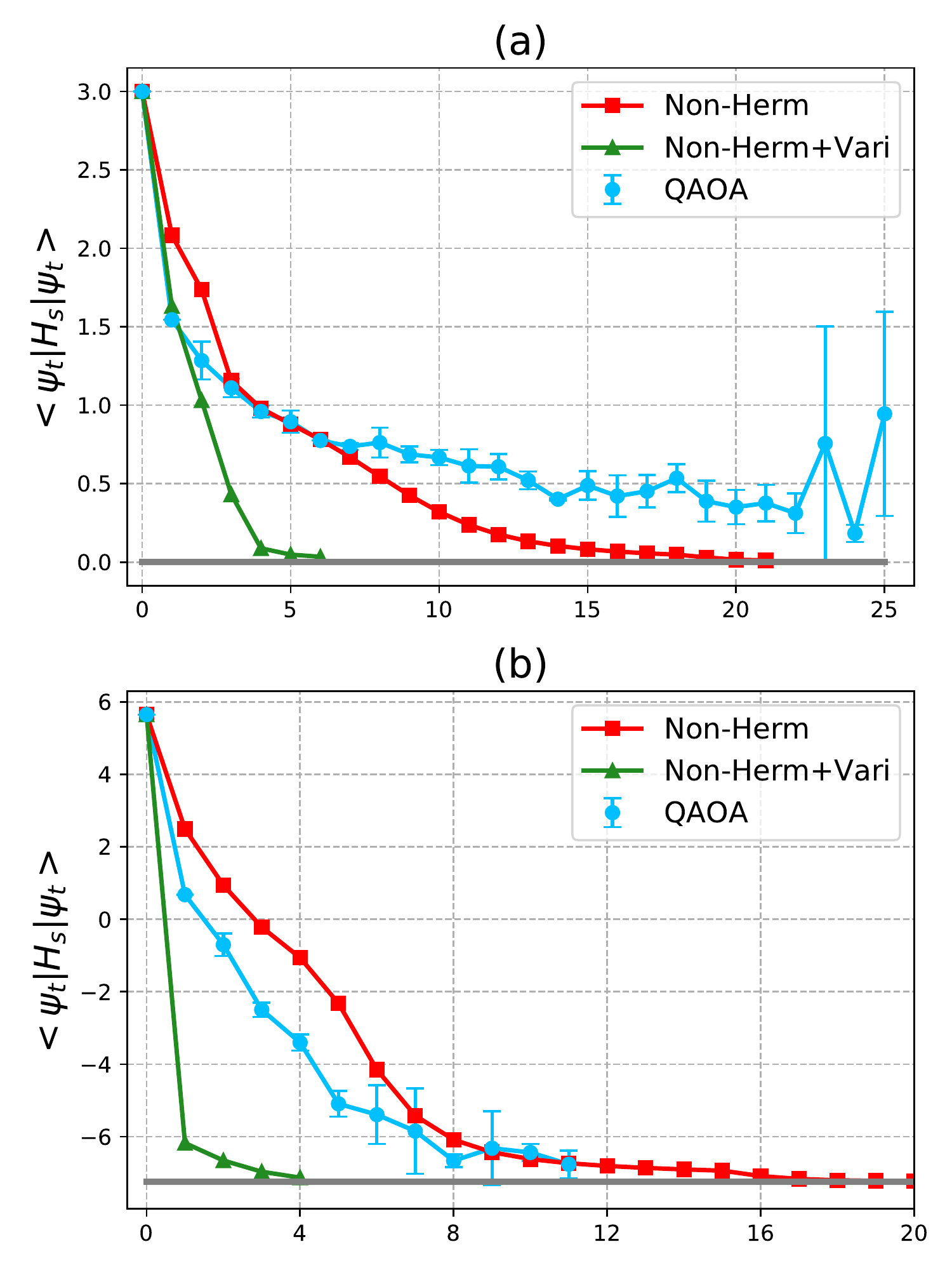}
\caption{The energy value obtained from different algorithm for different number of evolution steps. The error bar indicates the results from different initialization. (a) Simulation results for 3-SAT problem of  8 variables, details see Appendix~\ref{sec:3satmodels}. (b) Simulation results for 1D TFIM problem of  8 spins, details see Appendix~\ref{sec:TFIMmodels}. ``non-Herm" (Red) stands for non-Hermitian Hamiltonian simulation algorithm, ``non-Herm+Vari" (green) stands for non-Hermitian algorithm combined with variational blocks. ``QAOA" (blue) stands for standard QAOA algorithm with COBYLA optimizer. }
\label{fig:cos}
\end{figure}

We present the simulation results of the vanilla version of our non-Hermitian dynamical algorithm in Fig \ref{fig:cos} (red line). The results demonstrate that the non-Hermitian dynamical algorithm successfully prepares the ground state for both many-body quantum systems and classical combinatorial optimization problems.

\section{Hybrid variational blocks with non-Hermitian Hamiltonian simulation}
As discussed before, the non-Hermitian dynamical simulation (effectively implements the cosine filtering) gradually drives the initial quantum state towards the ground state. To accelerate this convergence and reduce circuit depth as much as possible, we propose to insert blocks of parameterized shallow quantum circuit between non-Hermitian propagators. As illustrated with the numerical examples below, the hybrid non-Hermitian-variational algorithm reaches the ground state at the same accuracy with a reduced circuit depth. Furthermore, we also compare the simulation efficiency of the hybrid non-Hermitian-variational algorithm with conventional QAOA, and show the superiority in circuit depth and measurement consumption for our method as elaborated in Appendix~\ref{sec:C}. 

The non-Hermitian-variational algorithm aims to combine the strength of the two distinct approaches: variational and Hamiltonian simulation methods.  On the one hand, variational methods have been actively developed to perform quantum simulation with shallow quantum circuits with attention to address the circuit structure design, quantum resource efficiency and barren plateau mitigation and so on ~\cite{sim2019expressibility, zhang2022variational, zhang2021variational}. On the other hand, the optimization of gate parameters in variational methods on a high-dimensional complex energy landscape is faced with challenges such as local minima and barren plateaus, which might be possibly overcome with the non-Hermitian propagation that is theoretically guaranteed to lower the overall energy of the quantum state.

We utilize several straightforward ideas to combine the two simulation techniques. The first one is to prepend a variational block in front of the non-Hermitian propagator. The variational module can be interpreted as improving the overlap of the initial state and the ground state; thus, it efficiently reduces the number of time steps needed by the following non-Hermitian propagation to reach the ground state. The benefits of adopting the variational module become clearer when the system size increases. As we know, the overlap $\Lambda$ between a randomly generated initial state and the ground state generally decays exponentially,  which potentially leads to a substantially increased non-Hermitian evolution time step to converge to the ground state. The detail on the correlation between the initial/final state overlap and the required number of non-Hermitian propagation time lengths is summarized in Appendix \ref{sec: A}. Beyond the simple idea, we further consider a hybrid simulation scheme by alternating the non-Hermitian Hamiltonian simulation blocks and variational blocks. We find that this hybrid scheme performs better than the former approach. 
Further details can be found in Appendix \ref{sec: B} and sketched in Fig. \ref{fig:circuit}. 
Because the variational modules are iteratively appended to the circuit (after each unit of non-Hermitian propagation), the variational modules can be optimized layer by layer to further reduce the consumption of quantum resources for optimization.

In Fig. \ref{fig:cos}, we first compare the numerical simulation of the pure non-Hermitian algorithm and the non-Hermitian-variational algorithm for the TFIM model and 3-SAT problems. The variational module we choose comprises only single-qubit gates. Hence, each variational module consists of an extremely shallow layer of quantum circuit. Yet we find that the non-Hermitian-variational algorithm substantially accelerates the ground-state preparation in comparison to the pure non-Hermitian algorithm. Since the non-Hermitian-variational algorithm adopts an alternating pattern of a simple variational module (single-qubit gates) and a dynamical propagator $U_{NH}=\exp(-i H_S\otimes Y_A dt)$ generated by the problem Hamiltonian, the circuit structure is actually highly comparable to the standard QAOA ansatz which also alternates between unitaries generated by single-qubit mixing Hamiltonian and the problem Hamiltonian. Thus, we also present numerical results given by the standard QAOA algorithm in Fig. \ref{fig:cos}. As shown in Fig. \ref{fig:cos}, for both many-body problem and combinatorial optimization problem, the non-Hermitian-variational algorithm outperforms QAOA with the same number of Hamiltonian blocks. A more detailed account on comparing the consumption of quantum resources for the two algorithms in terms of circuit depth and the number of measurement shots is given in Appendix \ref{sec:C}. The analysis indicates our proposed method is highly useful for non-trivial quantum simulations in the NISQ and the early fault-tolerant era.


The introduction of variational modules in the circuit substantially accelerates the convergence over purely non-Hermitian evolution in general. This acceleration implies we can deal with less number of simulation timesteps before reaching a satisfying convergence, and mitigate the problem of exponential difficulty in post-selection or post-processing. There is another inherent property we observe with the non-Hermitian-variational algorithm. Remarkably, the number of measurement shots needed for accurately estimating observable by post processing is greatly reduced with variational modules mixed. This is because the magnitude  of the denominator of Eq. \ref{eq:pprocessing} is greatly increased from zero. In short, the variational modules not only reduce the required duration of the non-Hermitian evolution to reach a desired ground state energy accuracy but also reduces the required number of measurement shots by improving the norm of the desired cosine filter component in the final state as detailed in Appendix \ref{sec:C}.


\section{Breaking the non-Hermitian evolution into pieces by state recording} 
To constrain the circuit depth, we consider further mitigating the issue of the exponentially diminishing success probability by breaking the full non-Hermitian evolution into several pieces. At every given number of time steps, we measure the time-evolved quantum state (i.e. conducting partial or full tomography to extract essential characteristics of this state) and approximately store the state with parameterized quantum circuits within a variational framework. Therefore, we can resume the non-Hermitian propagation with the variational state as the new initial state.

Similar ideas about state recording in a variational manner have been previously proposed in the literatures~\cite{lin2021real,barison2021efficient}.  Most of the previous methods attempted to capture the exact state while our strategy is more flexible and consumes fewer quantum resources on average as explained below. In addition to the full-state recording, we provide two customized reduced state-recording strategies that work well with classical combinatorial optimization problems and many-body quantum problems, respectively. 

\begin{figure}[htp]
\centering
\includegraphics[width=0.47\textwidth]{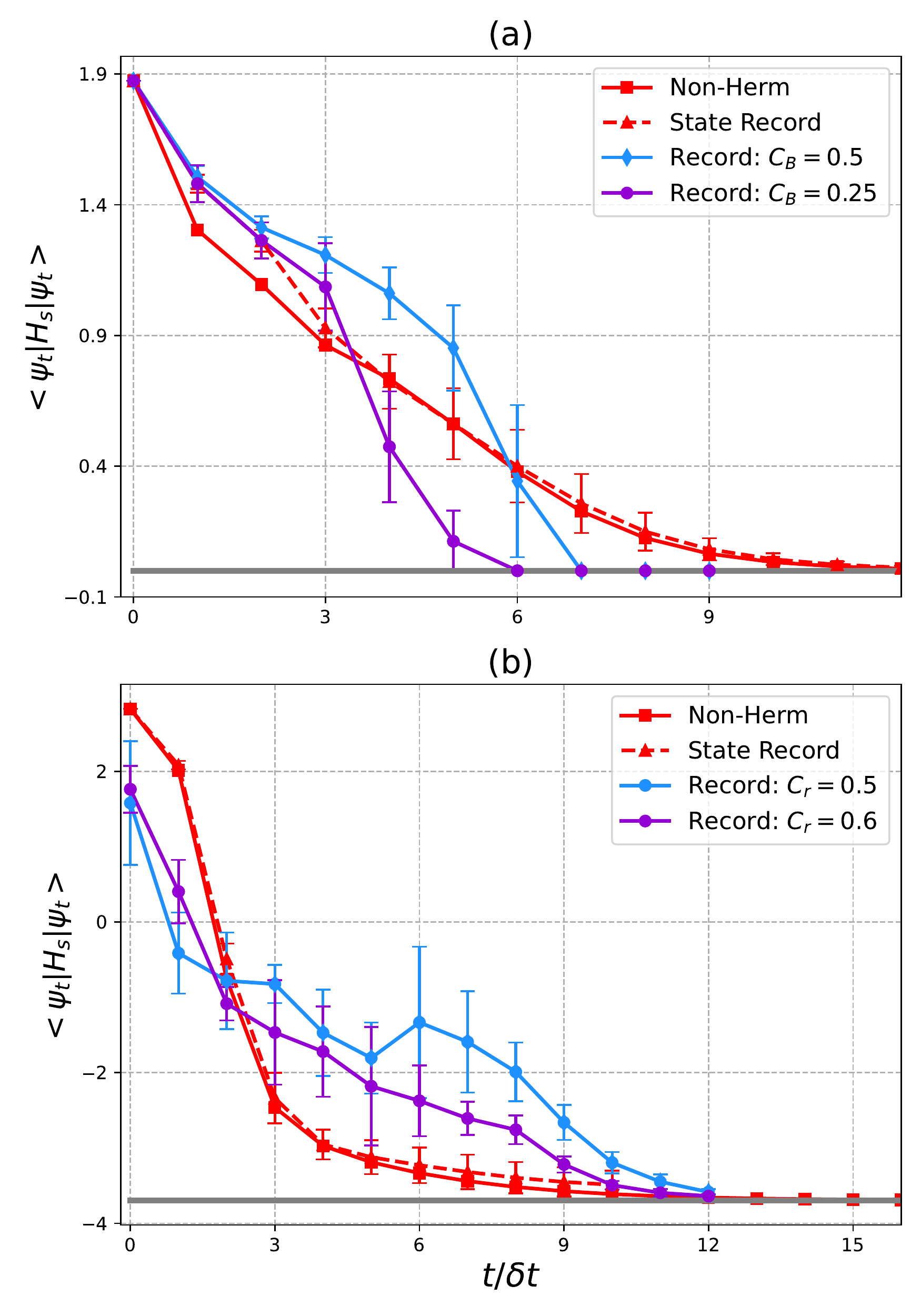}
\caption{The dynamic evolution of the system energy. We record on each time evolution step for demonstration. The error bar indicates the results from different initialization. (a) Simulation results of 3-SAT problem of  5 variables. ``non-Herm" (solid red line) for non-Hermitian algorithm, ``State Record" (dashed red line) for full recording of non-Hermitian evolution by variational ansatz. $C_B=0.5$ (blue) for reduced recording of non-Hermitian evolution by variational ansatz  with recording bound $C_B=0.5$. $C_B=0.25$ (purple) for reduced recording of non-Hermitian evolution by variational ansatz with recording bound $C_B=0.25$.
(b) Simulation results of TFIM problem of 4 spins. The variational ansatz is trained mostly on reduced recording for randomly chosen Pauli strings and followed by 3 steps of full recording. ``non-Herm" (solid red line) for non-Hermitian algorithm, ``State Record" (dashed red line) for fully recording of non-Hermitian evolution by variational ansatz. $C_r=0.5$ (blue) for reduced recording of non-Hermitian evolution by variational ansatz  with measurement ratio 0.5. $C_r=0.6$ (purple) for reduced recording of non-Hermitian evolution by variational ansatz with measurement ratio 0.6.    }
\label{fig:record}
\end{figure}

We break the non-Hermitian evolution at every $C$ steps to satisfy $\left \|\cos^C(H_S dt) \left|\psi_0 \right> \right \| > \eta$, where $\eta$ is a threshold parameter for suitable  number of measurement shots to arrive proper precision.  (details can be found in Appendix \ref{sec: A}). Namely, we apply $C$ consecutive blocks of $U_{NH}$ propagator to the initial state
\begin{equation}
U_{NH}^C=(e^{-iHs\otimes Y_A dt})^C.
\end{equation}
Let $V(\vec{\omega_m})$ be the parameterized circuit for
approximating the exact quantum state at the $m$-th step of the non-Hermitian evolution initialized with $\left|\psi_m\right\rangle=V(\vec{\omega_m}) \left|\psi_0\right>$. The parameters $\vec{\omega_m}\in \mathbb{R}^{p}$ denote the gate parameters in the quantum circuit $V(\vec{\omega_m})$ . The objective is to find an optimal $\vec{d\omega}$ to maximize the fidelity between the variational states $V(\vec{\omega}+\vec{d \omega})\ket{\psi_0}$ and $U_{NH}^CV(\vec{\omega})\ket{\psi_0}$,  
\begin{equation}\label{eq:obj-full}
\underset{\vec{d\omega} \in \mathbb{R}^{p}}{\arg \max }\left\|\left\langle\psi_0 \mid V^{\dagger}(\vec{\omega}+\vec{d\omega}) U^C_{NH}  V(\vec{\omega})\mid \psi_0\right\rangle\right\|^{2}.
\end{equation}
In practice, the state fidelity appearing in Eq.~(\ref{eq:obj-full}) can be evaluated with a quantum circuit by first creating the state $|\Psi_{\vec{\omega}}\>=V^{\dagger}(\vec{\omega}+\vec{d\omega}) U^C_{NH}  V(\vec{\omega})| \psi_0\>$ and then perform the following projective measurement,
\begin{equation} \label{eq:full record}
\frac{\<\Psi_{\vec{\omega}} \mid \mathcal{P_A}\mathcal{P_S}\mid \Psi_{\vec{\omega}}\>}{\<\Psi_{\vec{\omega}} \mid \mathcal{P_A}\mid \Psi_{\vec{\omega}}\>},
\end{equation}
where $\mathcal{P_A}=\frac{1}{2^{n_A}}(1+Z_1)(1+Z_2)\cdots (1+Z_{n_A})$ projects the ancilla qubit to the state $\left\vert 00\cdots0\right\rangle$, $\mathcal{P_S}=\frac{1}{2^{n_S}}(1+Z_1)(1+Z_2)\cdots (1+Z_{n_S})$ projects the system qubits to the state $|00\cdots 0\>$ (assuming $\vert \psi_0\rangle = |00\cdots 0\>$). $n_A=C$ is the cutoff length for the propagation (or the number of time steps), and $n_S$ denotes the number of system qubits. The derivation for Eq.~\ref{eq:full record} is delegated to Appendix \ref{sec:D}. We summarize how to simulate the non-Hermitian dynamics with the regular full-state recordings at every $C$-th step in Algorithm \ref{alg:full}, where $N_{rep}$ is the total number of state recordings performed. In Fig.~\ref{fig:record} we demonstrate successful ground state preparations for the TFIM and 3-SAT models by running Algorithm \ref{alg:full}. The specific variational ansatz used in these simulation experiments is provided in Appendix \ref{sec:D}.

\begin{algorithm}     
\caption{Concatenated non-Hermitian evolution with full-state recording in between}
\label{alg:full}
\begin{algorithmic}[1]
\INPUT $H_S$,  $dt$, $C$,$V(\vec{\omega})$,$N_{rep}$
\OUTPUT $E_f$
\STATE \textbf{Initialization} $\vec{\omega}=0$
\WHILE {$i \leq N_{rep}$}
\STATE Prepare variational quantum circuit:       $|\Psi_{\vec{\omega}}\>=V^{\dagger}(\vec{\omega}+\vec{d\omega}) U^C_{NH}   V(\vec{\omega})| \psi_0\>$, $U^C_{NH}=e^{(-iH_S\otimes Y_A dt)C}$.
\STATE Measure the quantum circuit in Z direction and calculate fidelity according to Eq.~\ref{eq:full record}.
\STATE Optimize $\vec{d \omega}$ based on gradient descent to maximize the fidelity according to Eq.~\ref{eq:obj-full}.
\STATE Update $\vec{\omega}=\vec{\omega}+\vec{d \omega}$.
\ENDWHILE
\STATE Return the final $\vec{\omega}_f$, calculate $E_f=\left< \psi_{\vec{\omega}_f} \mid H_S \mid \psi_{\vec{\omega}_f}  \right>$
\end{algorithmic}
\end{algorithm}

While the full-state recording is conceptually ideal, it consumes a substantial amount of quantum resources for measurements. Hence, we propose two other algorithms of reduced state recording for classical combinatorial optimization problem and many-body quantum problem, which requires fewer measurement operations.

For combinatorial optimization problems, inspired by a recent work called the recursive QAOA~\cite{bravyi2020obstacles}, we propose a reduced recording algorithm with cutoff. 
The method goes as follows. After each $C$ steps of non-Hermitian evolution, we measure  $M_{h_n}=\left\langle\psi^{*}\left|h_n\right| \psi^{*}\right\rangle$ on the instantaneous quantum state $\psi^{*}$ given by the non-Hermitian propagation, where $h_n$ is a Pauli string (such as $Z_i,Z_i Z_j,Z_iZ_kZ_j...$) that constitutes parts of the problem Hamiltonian for a combinatorial optimization problem. Then, we set all $M_{h_n}\geq C_B$ to be $1$, all $M_{h_n}\leq -C_B$ to be $-1$ and leave other $M_{h_n}$ untouched,
\begin{equation} \label{eq:extreme}
M_{h_n}'=\left\{\begin{array}{rl}-1, & \left(M_{h_n}\leq-C_B\right) \\ 1, & \left(M_{h_n}\geq C_B\right) \\ M_{h_n}, & \left(-C_B<M_{h_n}<C_B\right)\end{array}  \right. ,
\end{equation}
where $C_B$ is the bound we set for the confidence interval. The motivation to substitute $M_{h_n}'=\pm 1$ when $\vert M_{h_n}\vert > C_B$ follows the argument given by the recursive QAOA. To speed up in finding the ground state of $H_S$ by QAOA, recursive QAOA records the correlation properties of two spin observables presented in Hamiltonian $H_S$ to be correlated or anti-correlated according to the largest magnitude of measured correlation functions once optimizing the parameters to minimize $\left<\psi \mid H_S\mid \psi \right>$. Instead of using $M_{h_n}$ as in the full-state recording, the variational ansatz here is trained according to the updated value $M_{h_n}'$. We summarize the details of the reduced recording algorithm as Algorithm \ref{alg:reduce1} in Appendix \ref{sec:E}. The numerical experiments demonstrating the applicability of the reduced-recording algorithm on 3-SAT problems are presented in Fig. \ref{fig:record}(a) by setting $C_B=0.5$ or $C_B=0.25$. We find that the reduced-recording algorithm successfully simulates the desired ground state, and it is much more efficient than fully recording. 

Different from the recursive QAOA~\cite{bravyi2020obstacles}, we record the qubits' correlation statistics (measurement of bit strings in the computational basis) according to $C_B$ instead of simply updating a single $M_{h_n}$ with the largest absolute value and eliminating one qubit at each iteration. With (potentially) multiple $M_{h_n}$ values get updated in each iteration, as prescribed by Eq.~(\ref{eq:extreme}), we tremendously modify the objective function's landscape in the parameter space and allow more efficient search for optimal parameters in the examples we studied. The rationale behind the substitution $M^\prime_{h_n} = \pm 1$ is due to that the non-Hermitian algorithm always progressively filters out the high-energy components and converges towards the ground state. This core idea implies that the reduced recording can work as long as it captures the related part but not necessarily the full information of the original quantum state. Further analysis and justification for the reduced recording of $M_{h_n}$ are provided in Appendix \ref{sec:E}. In the appendix, we also show that the non-Hermitian algorithm with reduced recording may substantially outperform recursive QAOA with faster convergence, a point worth further study.


In terms of preparing the ground state for a many-body quantum Hamiltonian, we propose another reduced recording algorithm by randomly recording partial correlation statistics of the target state. This strategy decreases the number of measurements for the state recording. More specifically, after each $C$ steps of non-Hermitian evolution, we measure $M_{h_n}=\left\langle\psi^{*}\left|h_n\right| \psi^{*}\right\rangle$ with respect to the intermediate quantum state $\psi^{*}$ given by the non-Hermitian algorithm. Different from the full-state recording method, we do not consider every $h_n$ constituting the problem Hamiltonian. We randomly choose $N_r$ Pauli strings (from the problem Hamiltonian) for the state recording after every $C$-step non-Hermitian evolution, and we denote the ratio of the number of randomly selected Pauli strings to the total number of Pauli strings in $H_S$ by $C_r=N_r/N$. The optimization objective for this version of the reduced-state recording is to find an ansatz circuit that can reproduce these $M_{h_n}$ values as closely as possible.
\begin{equation}
\underset{\vec{\omega} \in \mathbb{R}^{p}}{\arg \min }\sum_{n=1}^{N_r}\left\|\left\langle\psi_0 \mid V^{\dagger}(\vec{\omega}) h_n V(\vec{\omega})\mid \psi_0\right\rangle-M_{h_n}\right\|^{2}.
\end{equation}

We can vary this sampling ratio $C_r$ along the course of the non-Hermitian evolution to ensure the final result correctly reflects the ground state behavior of the original Hamiltonian.
In the beginning, we can be more lenient and record an approximate state with variational ansatz at a smaller $C_r$ value. As the energy-filtering progress goes to the late stage, it is beneficial to more faithfully record the intermediate state $\psi^*$ by increasing $C_r$ to 1 gradually. Nevertheless, for the TFIM model we investigate in Fig. \ref{fig:record}(b), we find the reduced recording to work remarkably well even if we hold the sampling ratio $C_r$ at a fixed value much smaller than $1$ 
throughout the simulation except for the very last few steps we set $C_r = 1.0$. The details of this reduced recording algorithm for the quantum Hamiltonian are summarized as Algorithm \ref{alg:reduce2} in Appendix \ref{sec:E}.


\section{Discussion}
We propose a non-Hermitian Hamiltonian simulation algorithm to effectively realize an energy-filtering process that efficiently approximates the ground state on a quantum computer. The core subroutine of the proposed algorithm is Hamiltonian simulation algorithm, which is a well-established task with various efficient implementations and is known for potential quantum advantage. We further propose two techniques with variational toolbox, namely, 1) the hybridization with variational modules in circuit construction and 2) reduced variational state recording of intermediate evolved quantum states, to allow this non-Hermitian algorithm to work better with limited quantum resources in the NISQ and early fault-tolerant era. We demonstrate that our non-Hermitian algorithm can provide convincing performance boost for both combinatorial optimization problems and quantum many-body problems with much fewer quantum resources.

The hybridization with the variational module accelerates the energy-filtering and convergence towards the ground state. Specifically, we propose an algorithm that alternates the application of the non-Hermitian Hamiltonian evolution and variational transformation. The non-Hermitian-variational algorithm recovers the ground state more efficiently than the standard QAOA in our numerical studies. We remind that the non-Hermitian-variational algorithm also improves the success probability of post-selection and thus reduces the required number of measurements for post-processing the ancilla qubits. 

 We also propose a novel approach to break the full time evolution of the non-Hermitian Hamiltonian by variationally recording the intermediate quantum states with a parameterized quantum circuit. To further save quantum resources, we develop customized reduced recording strategies for combinatorial optimization problem and quantum many-body problem, respectively. By comparing the simulation time steps for convergence of full recording and reduced recording on the given problems, we conclude that the reduced recording algorithms can recover the ground states with less consumption of quantum resources. Specifically, the non-Hermitian Hamiltonian algorithm with reduced recording can consistently outperform QAOA for 3-SAT task under investigation. Also, the reduced recording method for many-body quantum problem provides a feasible approach to reduce the number of Pauli strings to be evaluated during the simulation, and make the algorithm more scalable and resource friendly. 
 
 Finally, we remind readers that all these variational gadgets we explored in this work can be utilized in many different quantum algorithm design scenarios as useful algorithm primitives. More importantly, the general idea to hybrid long-term quantum algorithms (with rigorous theoretical bounds on performance guarantees) and variational quantum algorithm pipelines and gadgets (requiring fewer quantum resources) is a very promising approach to achieve quantum advantage and will potentially play an important role in the NISQ and early fault-tolerant era.


\bibliographystyle{apsrev4-1}
\bibliography{ref}

\onecolumngrid

\appendix
\section{Non-Hermitian Hamiltonian evolution and cosine energy filtering} \label{sec: A}
In this section, we illustrate how the cosine energy filtering emerges from the non-Hermitian Hamiltonian evolution described in the main text. We will then summarize the key properties of cosine energy filtering for ground state simulation.

To simulate the ground state of a problem Hamiltonian $H_S$ in the Hilbert space of $n$ qubits, we prepare an initial state $\left|\psi_0\right>$. An ancilla qubit is then introduced and initially prepared in the state $\left|0\right>$. After undergoing a real-time evolution $U_{NH}=e^{-iHs\otimes Y_A dt}$ with timestep $dt$, the system-ancilla joint quantum state evolves to
\begin{equation}
\begin{array}{l}\left|\phi_{1}\right\rangle=U_{NH}\left|\psi_{0}\right\rangle|0\rangle \\ =\cos \left(H_{S} dt\right)\left|\psi_{0}\right\rangle|0\rangle-i \sin \left(H_{S} dt\right) \otimes Y_A\left|\psi_{0}\right\rangle|0\rangle \\ =\cos \left(H_{S} dt\right)\left|\psi_{0}\right\rangle|0\rangle+\sin \left(H_{S} dt\right)\left|\psi_{0}\right\rangle|1\rangle,
\end{array}
\end{equation}
where $Y_A$ is the usual Pauli operator applied on ancilla qubit.
By projecting the ancilla qubit to $\left|0\right>$, we obtain a quantum state in the system's Hilbert space, 
\begin{equation}
\left|\psi_1\right>=\frac{\cos\left(H_{S} dt\right)\left|\psi_{0}\right\rangle}{\| cos(H_S dt)\left|\psi_0\right> \|}.
\end{equation}
Repeating the real-time evolution block $U_{NH}$ $M$ times by recycling the ancilla qubit, we acquire final quantum state
\begin{equation}
\begin{array}{l}\left|\phi_{M}\right\rangle=U_{NH}^{M}\left|\psi_{0}\right\rangle|00 \ldots 0\rangle \\ =\cos ^{M}\left(H_{S} dt\right)\left|\psi_{0}\right\rangle|00 \ldots 0\rangle  +\cos ^{\mathrm{M}-1}\left(H_{S} dt\right) \cdot \sin \left(H_{S} dt\right)\left|\psi_{0}\right\rangle|00 \ldots 1\rangle+\cdots \\
+\cos\left(H_{S} dt\right) \cdot \sin^{M-1} \left(H_{S} dt\right)\left|\psi_{0}\right\rangle|01 \ldots 1\rangle+\sin ^{M}\left(H_{S} dt\right)\left|\psi_{0}\right\rangle|11 \ldots 1\rangle.
\end{array}
\end{equation}
We project $\left|\phi_M\right>$ to quantum state with ancilla qubit on state $\left|00\cdots 0\right>$, and finally arrive at a normalized state as the output 
\begin{equation}
\left|\psi_M\right>=\frac{\cos ^{M}\left(H_{S} dt\right)\left|\psi_{0}\right\rangle}{\| cos^M(H_S dt)\left|\psi_0\right> \|}.
\end{equation}
Thus the non-Hermitian Hamiltonian evolution effectively realizes the cosine energy filter. As discussed in the main text, the non-Hermitian Hamiltonian evolution can be naturally embedded as a straightforward real-time Hamiltonian evolution with the Hamiltonian $H_s \otimes Y_A$ in the joint Hilbert space involving an ancilla qubit. Since there are many well-established theoretical analyses and efficient implementations for Hamiltonian simulation $\exp(-i H t)$, our method could be super useful if we can efficiently deal with the post-selection or post-processing of ancilla qubits.

Aiming at preparing a good approximation of the ground state, the cosine-filter~\cite{ge2019faster} is a quantum simulation method that may outperform the phase estimation~\cite{kitaev1995quantum} based approach. As is known that the runtime of algorithms based on adiabatic algorithm~\cite{farhi2000quantum} depends inverse polynomially on the minimum spectral gap along the evolution path~\cite{jansen2007bounds}. And the probability of success for phase estimation is proportional to $\Lambda=\| \left< \psi_g \mid \psi_0\right>\|^2$, with $\left|\psi_0\right>$ the initial state and $\left|\psi_g\right>$ the ground state.
According to~\cite{ge2019faster}, the runtime of cosine-filter scales exponentially better in the allowed error to the real ground state, and polynomially better with the spectral gap and the overlap of the initialized trial state with the ground state. Specifically, since
\begin{equation}
\cos^M(H_S dt)\left|\psi_0\right>=\Lambda \cos^M(E_g dt)\left(\left|\psi_g\right>+\frac{\cos^M(H_s dt)}{\Lambda \cos^M(E_g dt)}\left|\psi_{\neg g} \right> \right),
\end{equation}
where $\left|\psi_{\neg g} \right>$ denotes excited states.
$\cos(x)$ is concave and decreasing in $\left[0,\pi/2 \right]$ and $\frac{\cos^M(H_S dt)}{\cos^M(E_g dt)}\left|\psi_{\neg g} \right> < e^{-\Omega(M(E_g dt)\Delta)}$. Where $\Delta$ is the minimum energy gap of $H_S$. According to~\cite{ge2019faster}, it is proved that high powers of $\cos(H_S dt)$ are approximately proportional to projectors onto the ground state, 
\begin{equation}
\left\| \frac{\cos ^{M}\left(H_{S} d t\right) \left\vert \psi_{0}\right\rangle}{\| \cos ^{M}\left(H_{S} d t\right) \left\vert \psi_{0}\right\rangle \|}-\left \vert \psi_{g}\right\rangle \right\| =O(\epsilon),
\end{equation}
where the required number of $M$ timesteps to get $\epsilon$-close to $\left| \psi_g  \right>$  is given by
\begin{equation} \label{eq:M}
M= \Theta \left( \frac{1}{\Delta^2}\log^2{\frac{1}{\chi^{\epsilon}}} \right),
\end{equation}
here $\chi$ is the lower bound of initial state/ground state  overlap $\Lambda$.
The norm of quantum state with $M$ non-Hermitian blocks reads as,
\begin{equation}
\label{a8}
\left\| \cos^M(H_S dt) \left| \psi_0 \right> \right\|=\Omega(\Lambda). 
\end{equation}

There are two perspectives in terms of energy estimation. For post-selection perspective, we must ensure that all M ancilla qubits are post-selected to state $\vert 0\rangle$, leaving the final system state in $\vert \psi_M\rangle$, and we then can measure the Hamiltonian expectation on this output state. The other post-processing perspective doesn't rely on post-selection, instead we regard the final state jointly defined on the system qubits and M ancilla qubits as $\vert \phi_M \rangle$ and then we estimate the expectation for $H_S\mathcal{P}=H_S(1+Z_1)(1+Z_2)\cdots (1+Z_M)$, where exponential Pauli string terms are required to measure na\"ively.

The complexity obstruction applies to both perspectives. In terms of the post-processing calculation of energy expectation in the main text
\begin{equation}
E_{M}=\frac{\left\langle\phi_{M}\left|H_{S} \mathcal{P}\right| \phi_{M}\right\rangle}{\left\langle\phi_{M}|\mathcal{P}| \phi_{M}\right\rangle},
\end{equation} 
the number of measurement shots needed for the results to reach a certain accuracy is inversely proportional to $\Lambda$ as the magnitude of the dominator is given by Eq. \ref{a8}. Equivalently, from post-processing perspective, the number of experiments required for post-select ancilla qubits to succeed is also inversely proportional to $\Lambda$, which is in general exponentially small in general.
Under an extremely weak assumption that there is a random initial state for $n$-qubit system,  the lower bound on $\Lambda$ is $e^{-O(\log{2^n})}$. As a result, for non-Hermitian algorithm, the number of  measurement shots grows exponentially as the scaling up of system size for random initial state. 

As discussed in the main text, instead of doing a generic post-selection on ancilla qubits with a cumulative success probability that drops exponentially, we practically choose the classical post-processing instead as given in Eq.~(\ref{eq:processing}). In this perspective, we can do some simplifications by taking advantage of the fact that all ancilla qubits essentially have identical effects on the system evolution due to the time
translational invariance. 

Firstly, note that the projector satisfies $\mathcal{PP=P}$ and $\left[\mathcal{P},H_S\right]=0$, we can then prove that
\begin{equation}
\begin{array} {ccc}
&  \left\langle\phi_{1}\left|Z_1\right| \phi_{1}\right\rangle=\left\langle\phi_{2}\left|Z_1I_2\right| \phi_{2}\right\rangle =\left\langle\phi_{3}\left|Z_1I_2I_3\right| \phi_{3}\right\rangle~\cdots \\
&  \left\langle\phi_{1}\left|H_SZ_1\right| \phi_{1}\right\rangle=\left\langle\phi_{2}\left|H_SZ_1I_2\right| \phi_{2}\right\rangle =\left\langle\phi_{3}\left|H_SZ_1I_2I_3\right| \phi_{3}\right\rangle~\cdots .
\end{array}
\end{equation}
And since $\left\langle\phi_{M}\left|Z_i\right| \phi_{M}\right\rangle = \left\langle\phi_{0}\left|e^{iH_S\otimes Y_i dt}Z_i e^{-iH_S\otimes Y_i dt}\right| \phi_{0}\right\rangle$, we have
\begin{equation}
\begin{array} {cccc}
&  \left\langle\phi_{M}\left|Z_1\right| \phi_{M}\right\rangle=\left\langle\phi_{M}\left|Z_2\right| \phi_{M}\right\rangle =\left\langle\phi_{M}\left|Z_3\right| \phi_{M}\right\rangle~\cdots \\
&  \left\langle\phi_{M}\left|Z_1Z_2\right| \phi_{M}\right\rangle=\left\langle\phi_{M}\left|Z_1Z_3\right| \phi_{M}\right\rangle =\left\langle\phi_{M}\left|Z_2Z_3\right| \phi_{M}\right\rangle~\cdots  \\
&  \cdots \\
& \left\langle\phi_{M}\left|Z_1Z_2\cdots Z_n\right| \phi_{M}\right\rangle
\end{array}
\end{equation}
Since $\left\langle\phi_{M}\left|H_SZ_i\right| \phi_{M}\right\rangle=\left\langle\phi_{0}\left|e^{iH_S\otimes Y_i dt}H_S Z_i e^{-i H_S\otimes Y_i dt}\right| \phi_{0}\right\rangle$, we have
\begin{equation}
\begin{array} {cccc}
& \left\langle\phi_{n}\left|H_SZ_1\right| \phi_{n}\right\rangle=\left\langle\phi_{n}\left|H_SZ_2\right| \phi_{n}\right\rangle =\left\langle\phi_{n}\left|H_SZ_3\right| \phi_{n}\right\rangle~\cdots \\
& \left\langle\phi_{n}\left|H_SZ_1Z_2\right| \psi_{n}\right\rangle=\left\langle\phi_{n}\left|H_SZ_1Z_3\right| \phi_{n}\right\rangle =\left\langle\phi_{n}\left|H_SZ_2Z_3\right| \phi_{n}\right\rangle~\cdots  \\
& \cdots \\
& \left\langle\phi_{n}\left|H_SZ_1Z_2\cdots Z_n\right| \phi_{n}\right\rangle
\end{array}
\end{equation}
Thus,we simplify the estimation on energy expectation from post-processing perspective as:
\begin{equation}
E_{M}=\frac{\mathcal{C}_M^0\left\langle\phi_{0}\left|H_{S}\right| \phi_{0}\right\rangle+\sum_{i=1}^M \mathcal{C}_M^i \left\langle\phi_{i}\left|H_{S}Z_1Z_2\cdots Z_i\right| \phi_{i}\right\rangle }{\mathcal{C}_M^0\left\langle\phi_{0}\vert \phi_{0}\right\rangle+\sum_{i=1}^M \mathcal{C}_M^i \left\langle\phi_{i}\left|Z_1Z_2\cdots Z_i\right| \phi_{i}\right\rangle}.
\end{equation}

\section{Hybrid approach by interspersing non-Hermitian evolution with variationally tuned adjustments } \label{sec: B}
In the NISQ era and early fault-tolerant period, we acknowledge that the quantum circuit should be kept to a minimal depth if possible. Hence, we propose to combine the non-Hermitian algorithm with variational method to accelerate the simulation towards the ground state. Since a parameterized quantum circuit may possess strong expressibility and approximate many physically relevant quantum state~\cite{sim2019expressibility} with shallow depth, we believe that the addition of variational modules may speed up the simulation process and provide a better solution with shallower circuit depth.

 On the other hand, the non-Hermitian Hamiltonian evolution algorithm with rigorous guarantee can help variational optimization on a complex energy landscape, that usually suffers from trapping in local minima and barren plateaus. The energy filtering process works with different mathematical machinery to drive the system towards lower energy. Therefore, the non-Hermitian evolution can help get a quantum state out of local minimum or barren plateau common in pure variational optimization setup.
\begin{figure}[htp]
\centering
\includegraphics[width=0.5\textwidth]{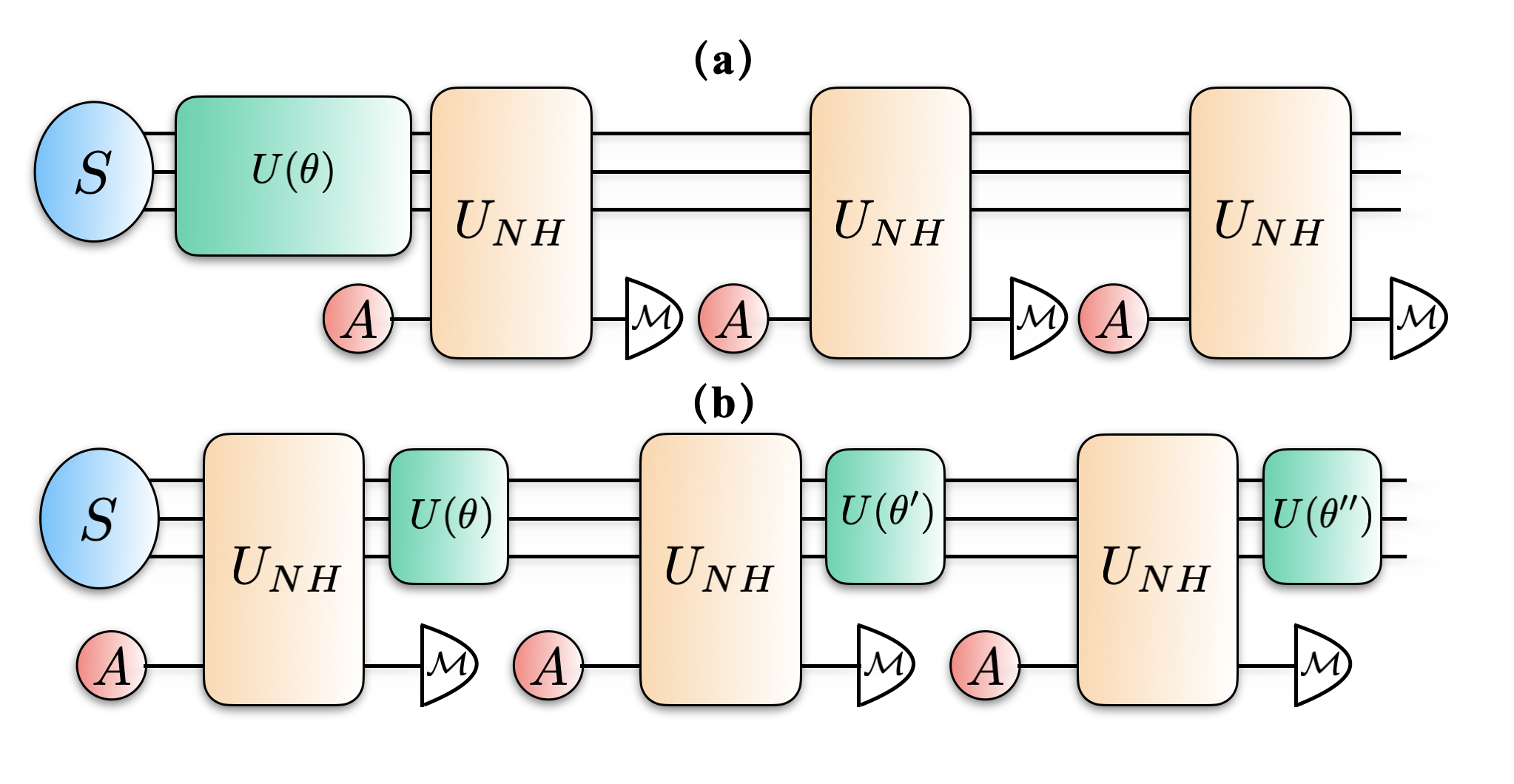}
\caption{The quantum circuit for non-Hermitian algorithm combined with variational block. (a) The circuit structure of applying a variational block in front of multiple non-Hermitian blocks. (b) The circuit structure of alternatively applies non-Hermitian blocks and variational blocks.  }
\label{fig:cucu}
\end{figure}

\begin{figure}[htp]
\centering
\includegraphics[width=0.85\textwidth]{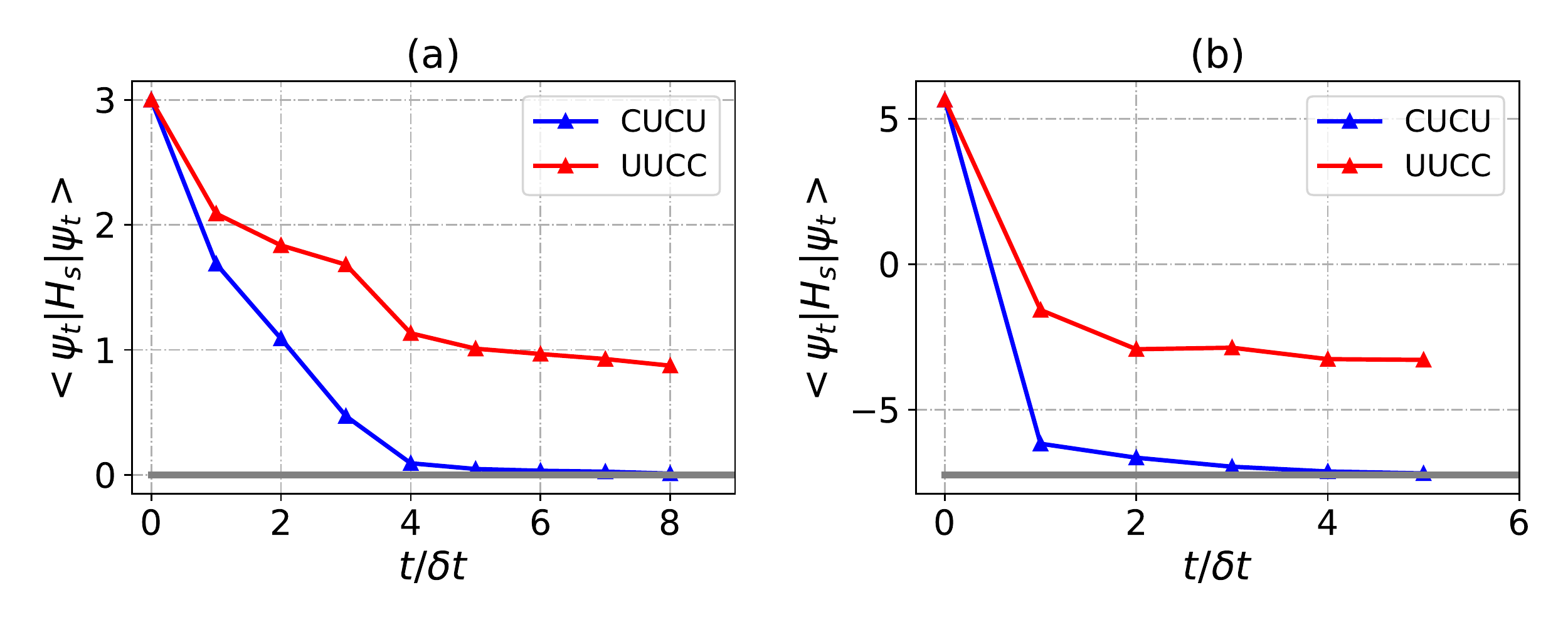}
\caption{The simulation results of non-Hermitian algorithm combined with variational blocks. (a) Results for 3SAT of 8 variables. (b) Results for TFIM of 8 spins. ``CUCU" (blue) denotes results of variational then non-Hermitian circuit  structure. ``UUCC" (red) denotes the results of interspersing non-Hermitian evolution with variational adjustments.  }
\label{fig:cucuexp}
\end{figure}

We compare two common hybrid quantum structures that combine the non-Hermitian algorithm and the variational method, see Fig.~\ref{fig:cucu}. One intuitive idea is to apply a variational ansatz layer in front of the non-Hermitian propagator as displayed in Fig.~\ref{fig:cucu}(a). In this case, an obvious objective for the ansatz layer is to improve the overlap of initial state and target state, thus we can efficiently reduce the Hamiltonian evolution time steps required by subsequent non-Hermitian evolution, see Eq.~\ref{eq:M}. 
The second approach we explore is to alternate the application of non-Hermitian block and variational block as displayed in Fig.~\ref{fig:cucu}(b). We investigate the effects of these two approaches by conducting numerical experiments on 3-SAT and TFIM examples, as given in Fig.~\ref{fig:cucuexp}. Here ``UUCC" denotes the method of applying variational ansatz before the non-Hermitian evolution, ``CUCU" denotes the method of interspersing the short-time non-Hermitian propagators with variational blocks. 
In experiments reported in Fig.~\ref{fig:cucuexp}, we adopt the simplest possible variational circuit in ``CUCU" layout, namely, we only put one layer of single-qubit gates in each variational block).  For ``UUCC" we use much more complex ansatz layout, which consists of not only layers of single qubit gates but also entangling gates. Hence, a typical ansatz structure in the ``UUCC" scheme costs more quantum gates than the number of single-qubit gates contained in the variational blocks in the ``CUCU" scheme. As shown in the figure, the second method (``CUCU") delivers better results for both 3-SAT problem of 8 variables and a TFIM of 8 spins even it has more light-weighted variational blocks.  Thus, we choose it as 
the standard layout to combine non-Hermitian algorithm and variational method to illustrate the power of non-Hermitian-variational algorithm in the main text.

\section{Resource consumption estimation} \label{sec:C}

\begin{figure}[htp]
\centering
\includegraphics[width=0.9\textwidth]{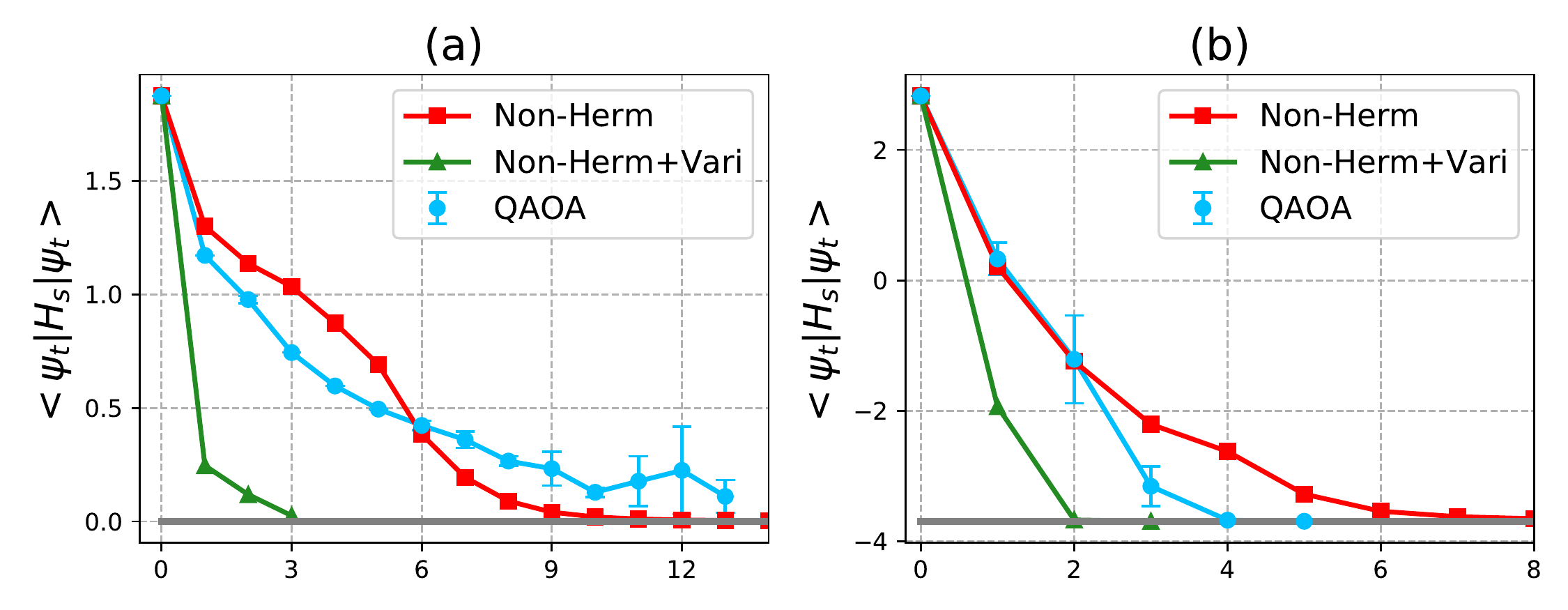}
\caption{The ground state energy estimation with different blocks of quantum circuit primitives. (a) Simulation results for 3-SAT problem of  5 variables.  (b) Simulation results for 1D TFIM problem of  4 spins. ``non-Herm" (Red) denotes the result of non-Hermitian algorithm, ``non-Herm+Vari" (green) denotes the result of non-Hermitian algorithm combined with variational block. ``QAOA" (blue) denotes the result of QAOA algorithm.}
\label{fig:eff2}
\end{figure}

In the main text we conclude that the non-Hermitian algorithm combined with variational method (non-Hermitian-Variational algorithm) is capable to boost the ground state simulation for both quantum many-body problems and classical combinatorial optimization problems compared to the original non-Hermitian algorithm and conventional QAOA algorithm. Here, we provide further results with different system sizes on the ground state preparation for 3-SAT problem and TFIM problem, see Fig.~\ref{fig:eff2}. The results indicate that the advantage of non-Hermitian-Variational algorithm is universal and doesn't rely on the specific details of the tasks.

We now analyze the resource consumption of non-Hermitian (non-Hermitian-Variational) algorithm and QAOA in terms of the depth of quantum circuit $N_{depth}$ and the number of measurements for convergence $N_{meas}$.

To perform the non-Hermitian (non-Hermitian-Variational) or QAOA experiments, the circuit depth $N_{depth}$ is defined based on the Hamiltonian evolution steps $N_{step}$ needed for the algorithm to converge to an output state with energy within a certain percentage (e.g. $ 1\%$) of the exact ground state energy  and the circuit depth of a single evolution step $N_U$, 
\begin{equation} \label{eq:depth}
N_{depth}=N_{step}\times N_U.
\end{equation}
For QAOA simulation, $N_U$ is decided by the circuit depth of realizing  $e^{iH_S\beta}$ and $e^{i H_B \gamma}$ from Trotterization. For non-Hermitian algorithm, $N_U$ is decided by the circuit depth of realizing $e^{-i H_S \otimes Y_A dt}$. For non-Hermitian-Variational algorithm, $N_U$ is determined by the circuit depth of realizing $e^{-i H_S \otimes Y_A dt}$ together with the circuit depth of an additional variational block.

\begin{figure}[htp]
\centering
\includegraphics[width=0.5\textwidth]{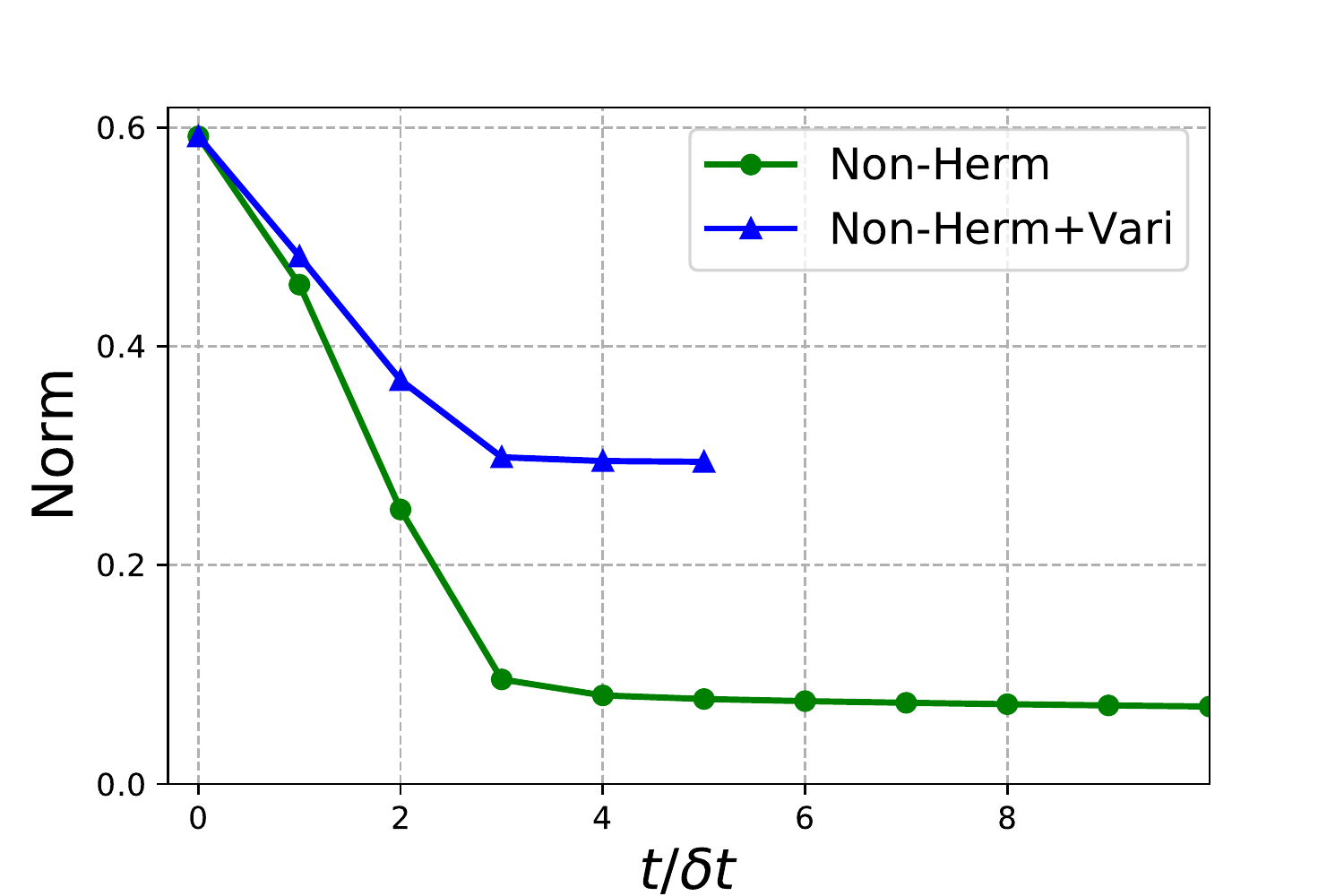}
\caption{The norm of quantum state calculated along with evolution for 3-SAT problem of 8 variables. ``non-Herm" (green) presents the result of non-Hermitian algorithm.  ``non-Herm+vari" (blue) presents the result of non-Hermitian-Variational algorithm. The introduction of the variational blocks greatly mitigates the ``sign problem" for the post-processing. }
\label{fig:norm}
\end{figure}

The number of measurement $N_{meas}$ required to conduct the simulation is determined by the number of Pauli basis that required to estimate all Pauli string terms presented in the Hamiltonian $N_h$, the number of parameters to be optimized $N_{para}$, ($2 N_{step}$ in QAOA case), estimation of average iterations $N_{ite}$ for the optimization of a single parameter by a classical optimizer, and measurement shots $N_{shots}$ needed for a quantum circuit to estimate the observable up to given accuracy,
\begin{equation} \label{eq:meas}
N_{meas}=N_h  \times N_{para}\times N_{ite} \times N_{shots}.
\end{equation}
For non-Hermitian algorithm without variational block, we set $N_{para}=N_{ite}=1$.  In non-Hermitian-Variational algorithm, we optimize the parameters of variational block in a block-by-block fashion, thus the number of parameters to be optimized is constrained along the evolution and can effectively avoid local minima. According to Appendix \ref{sec: A}, the measurement shots of non-Hermitian algorithm to calculate the results up to a certain accuracy is inversely proportional to $\Lambda$,  where $\Lambda$ is the overlap between initial state and ground state. 
From the simulation results, the non-Hermitian-Variational algorithms largely reduce the evolution steps required by the original non-Hermitian algorithm and demand much less number of measurement shots owing to the improvement on the norm of final quantum state, see Fig.~\ref{fig:norm}.

We focus on the examples studied in this work, $4$-site/$8$-site 1D TFIM and $5$-variable/$8$-variable 3-SAT problems to present the quantitative resources required for non-Hermitian-Variational algorithm and QAOA in Table~\ref{tab:resource}. For circuit depth estimation, the data presented in the table following Eq.~\ref{eq:depth}, e.g. $4*15 = 60$ means $N_{step}=4,N_U=15$. The different $N_{step}$ for different algorithms ensure that the final accuracy for ground state energy is the same. In terms of measurement shots estimation, the data presented in the table following Eq.~\ref{eq:meas}, e.g. $5*(4*2) = 40$ means $N_h=5, N_{para}=4*2$, $4$ is the evolution steps $N_{step}$. We ignore $N_{ite}$ and $N_{shots}$ here as they are similar for QAOA and non-Hermitian-variational algorithm. $N_{para} = 2 $ for "non-Hermi+Vari" in case "TFIM-4" is because the variational module considered in an evolution step only including two layers of single-qubit rotations $Rz(\theta)Rx(\theta')$, and all single-qubit gates in one layer share the same parameters. $N_{para} = 3 $ for "non-Hermi+Vari" in case "TFIM-8" is because the variational module considered here only including there layers of single-qubit rotations $Rz(\theta)Rx(\theta')Rz(\theta'')$, and again, all qubits share the same parameters. $N_{para} = 15 $ for "non-Hermi+Vari" in case "3-SAT-5" is because the variational module considered here including there layers of single-qubit rotations $Rz(\theta)Rx(\theta')Rz(\theta'')$, and all gates have different parameters. $N_{para} = 12 $ for "non-Hermi+Vari" in case "3-SAT-8" is because the variational module considered here including there layers of single-qubit rotations $Rz(\theta)Rx(\theta')Rz(\theta'')$, and half of the qubits share the same parameters in an even-odd fashion. 
\begin{table}  \label{tab:resource}
\centering
\caption{Quantum Resource Consumption for Different Algorithms}  
\begin{tabular*}{12cm}{ccccc}  
\hline  
\     & TFIM-4  & TFIM-8 & 3-SAT-5 & 3-SAT-8 \\  
\hline  
QAOA(dep.)           & 4*15 = 60 & 13*27 = 351 & 15*69 = 1035 & 28*1774 = 49672 \\  
non-Herm+Vari(dep.)  & 2*42 = 84 & 4*82 = 328 & 3*132 = 396 & 6*2307 = 13842 \\  
QAOA(meas.)          & 5*(4*2) = 40 & 9*(13*2) = 234 & 1*(15*2) = 30 & 1*(28*2)= 56 \\  
non-Herm+Vari(meas.)          & 5*2 = 10 & 9*3 = 27 & 1*15 =15 & 1*12 = 12 \\  
\hline  
\end{tabular*}  
\end{table}

\section{Cut off non-Hermitian algorithm by state recording} \label{sec:D}

Although non-Hermitian-variational algorithm is an efficient method to accelerate ground state simulation and significantly suppress the circuit depth, it is worth noting that the introduction of variational module breaks the time translation invariance leading to Eq.~(\ref{eq:pprocessing}). 
To further suppress the exponential cost of the algorithm, here we propose an approach to break the continuity of the non-Hermitian process (or non-Hermitian-Variational process) by recording the intermediate quantum state with a variational ansatz stored in the parameterized circuit after every $C$ steps (as implied by $U_{NH}^C$ in the following equations in this section).  
Without loss of generality, we assume $U_{NH}^C$ containing variational blocks as given in the `UCUC' scheme (introduced in Appendix~\ref{sec:C}). The idea of breaking the non-Hermitian process into small pieces helps to avoid a long evolution time that goes beyond the coherence time of the current generation of hardware and renders the post-selection success probability exponentially small. Particularly, we break the non-Hermitian evolution process into $C$ steps, which satisfies 
\begin{equation}
\left \|\cos^C(H_S dt) \left|\psi_0 \right> \right \| > \eta,
\end{equation}
where $\eta$ is a bound set for a target number of measurement shots. 

The objective function for full recording of a quantum state reads,
\begin{equation}
\underset{d w \in \mathbb{R}^{p}}{\arg \max }\left\|\left\langle\psi_0 \mid V^{\dagger}(\vec{\omega}+\vec{d\omega}) U^C_{NH}  V(\vec{\omega})\mid \psi_0\right\rangle\right\|^{2}.
\end{equation}
where $V$ is the PQC storing the intermediate quantum state.
In practice, this full recording of quantum state can be realized by firstly building a quantum circuit as  $|\Psi_{\omega}\>=V^{\dagger}(\omega+d\omega) U^C_{NH}  V(\omega)| \psi_0\>$. The interpretation of this circuit is as following. First, the quantum circuit $V(\omega)$ properly constructs the recorded state from the previous segment of $C$-steps evolution (i.e. the state $V(\omega)\left|\psi_0\right>$), then we execute the next $C$-steps of non-Hermitian-Variational algorithm using the quantum circuit $U^C_{NH}$. Finally, we want to represent this C-steps evolved state with the same variational circuit but having updated parameters as implied by $V^{\dagger}(\omega+d\omega)$.  This objective function can be evaluated by conducting projective measurements of $\mathcal{P_A}=\frac{1}{2^{n_A}}(1+Z_1)(1+Z_2)\dots (1+Z_{n_A})$ to post-process the ancilla qubit into state $\left|00\cdots0\right>$ and  $\mathcal{P_S}=\frac{1}{2^{n_S}}(1+Z_1)(1+Z_2)...(1+Z_{n_S})$ to realize the measurement on system operator $\hat{O}=\left|0\right>\left<0\right|$, 
\begin{equation} 
\underset{d w \in \mathbb{R}^{p}}{\arg \max }\frac{\<\Psi_{\omega} \mid \mathcal{P_A}\mathcal{P_S}\mid \Psi_{\omega}\>}{\<\Psi_{\omega} \mid \mathcal{P_A}\mid \Psi_{\omega}\>}.
\end{equation}
To avoid barren plateaus for the global loss function, local loss function that pins every qubit into $\vert 0\rangle$ state can also be exploited.

\begin{figure}[htp]
\centering
\includegraphics[width=0.7\textwidth]{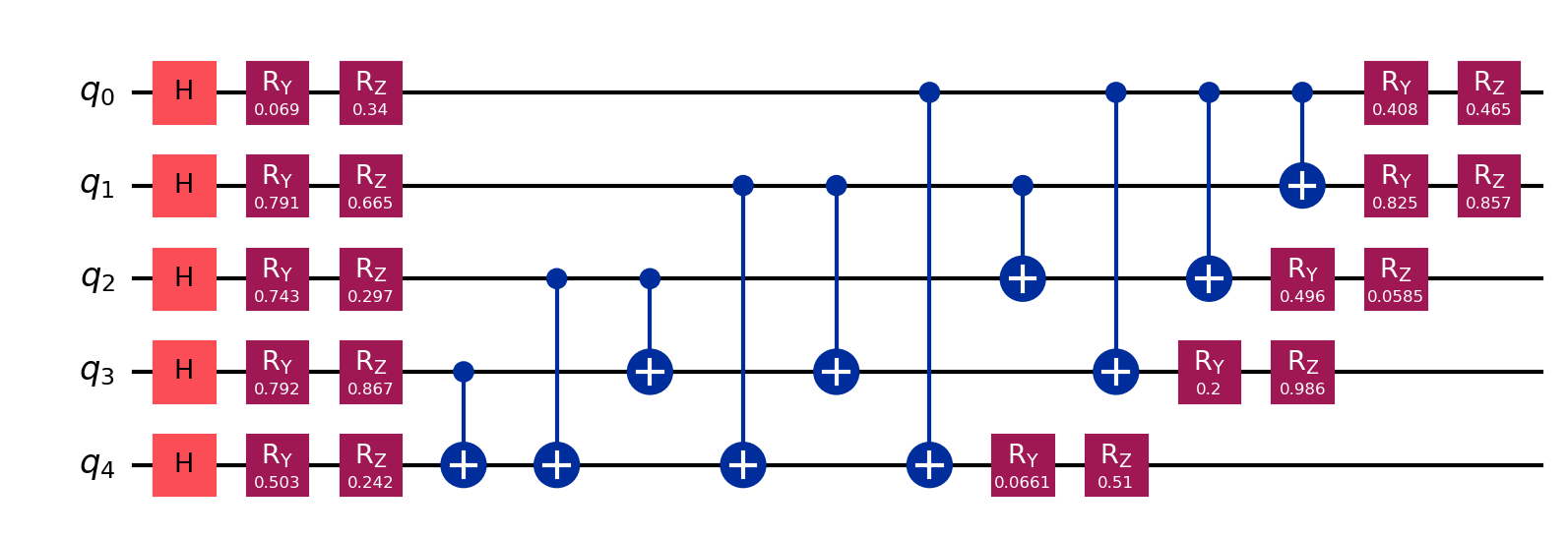}
\caption{The variational ansatz used for state recording.}
\label{fig:ansatz}
\end{figure}

The quantum variational ansatz $V_{\omega}$ used in the main text to record the quantum state output by the non-Hermitian algorithm is shown in Fig.~\ref{fig:ansatz}.  The Hadamard gates are applied firstly to realize the initial quantum state of a uniform superposition of all computational basis state,  $\left|\psi_0\right>=\frac{1}{\sqrt{2^n}}\left|++\cdots +\right>$. 

\section{Reduced state recording} \label{sec:E}
In the main text, in order to further reduce the resource consumption of state recording, we propose two bespoke methods as reduced recording for classical combinatorial optimization problem (CO) and quantum many-body problem, respectively. In this section, we present the algorithms of reduced state recording in Algorithm \ref{alg:reduce1} and Algorithm \ref{alg:reduce2}. 

\begin{algorithm} 
\caption{Reduced state recording for CO problem}
\label{alg:reduce1}
\begin{algorithmic}
\INPUT $H_S$, $dt$, $C$,$V(\vec{\omega})$,$N_{rep}$,$C_B$
\OUTPUT $E_f$
\STATE \textbf{Initialization} $\vec{\omega}=0$
\WHILE {$i \leq N_{rep}$}
\STATE Prepare variational circuits combined of two parts: $V(\vec{\omega})$, $U_{NH}^C$. $U_{NH}=e^{-iH_S\otimes Y_A dt}$. 
\STATE Obtain state $\left|\psi^{*}\right> = U_{NH}^C \vert \psi_0\rangle$ and $\vert \psi(\vec{\omega})\rangle = V(\vec{\omega})\vert \psi_0\rangle$.
\STATE Measure $M_{h_n}=\left\langle\psi^{*}\left|h_n\right| \psi^{*}\right\rangle$ for $n=1\cdots N$. Revise $M_{h_n}$ to $M'_{h_n}$ according to Eq.~\ref{eq:extreme} given $C_B$.
\STATE Measure $M^{\omega}_{h_n}=\left\langle\psi(\vec{\omega})\left|h_n\right| \psi(\vec{\omega})\right\rangle$. 
\STATE Optimize $\vec{\omega}$ based on gradient descent to minimize $\sum_{n=1}^N{\|M^{\omega}_{h_n}-M'_{h_n}\|}$
\ENDWHILE
\STATE Return the final $\vec{\omega}_f$,  calculate $E_f=\langle \psi({\vec{\omega}}_f) \vert H_S \vert \psi({\vec{\omega}}_f)\rangle$.
\end{algorithmic}
\end{algorithm}

\begin{algorithm} 
    \caption{Reduced state recording for quantum many-body problem}
    \label{alg:reduce2}
  \begin{algorithmic}
    \INPUT $H_S$,  $dt$, $C$,$V(\vec{\omega})$,$N_{rep}$,$N_r$
    \OUTPUT $E_f$
      \STATE \textbf{Initialization} $\vec{\omega}=0$
      \WHILE {$i \leq N_{rep}$}
      \STATE Prepare variational circuits combined of two parts: $V(\vec{\omega})$, $U_{NH}^C$. $U_{NH}=e^{-iH_S\otimes Y_A dt}$. 
\STATE Obtain state $\left|\psi^{*}\right> = U_{NH}^C \vert \psi_0\rangle$ and $\vert \psi(\vec{\omega})\rangle = V(\vec{\omega})\vert \psi_0\rangle$.
\STATE Randomly sampling $N_r$ Pauli string operators from $H_S$.
      \STATE  Measure $M'_{h_n}=\left\langle\psi^{*}\left|h_n\right| \psi^{*}\right\rangle$ for $n=1,2...N_r$.
      \STATE Measure $M^{\omega}_{h_n}=\left\langle\psi(\vec{\omega})\left|h_n\right| \psi(\vec{\omega})\right\rangle$.
      \STATE Optimize $\vec{\omega}$ based on gradient descent to minimize $\sum_{n=1}^{N_r}{\|M^{\omega}_{h_n}-M'_{h_n}\|}$
      \ENDWHILE
  \STATE Return the final $\vec{\omega}_f$,  calculate $E_f=\langle \psi({\vec{\omega}}_f) \vert H_S \vert \psi({\vec{\omega}}_f)\rangle$.
    \end{algorithmic}
\end{algorithm}

\begin{figure}[htp]
\centering
\includegraphics[width=0.9\textwidth]{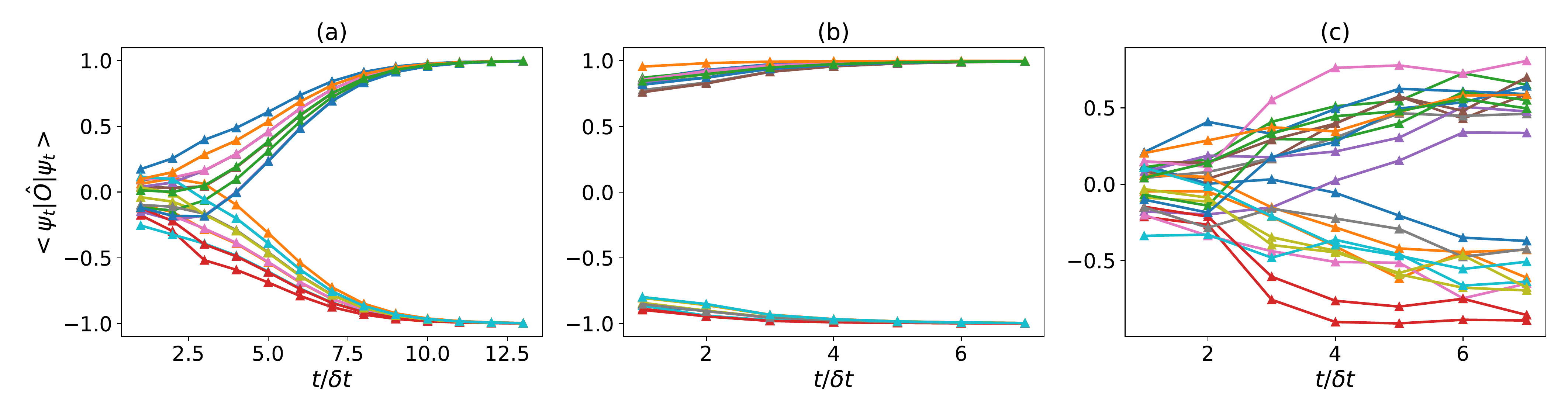}
\caption{The expectation value of Pauli strings $\hat{O}$ in 3-SAT $H_S$ calculated with different Hamiltonian evolution steps. (a) Results of non-Hermitian algorithm; (b) Results of non-Hermitian-Variational algorithm; (c) Results of QAOA.}
\label{fig:co}
\end{figure}

In the main text, we present the simulation results of reduced recording for the combinatorial optimization problem. In Fig.~\ref{fig:co} we take 3-SAT problem as an example to present how the expected values  $M_{h_n}=\left\langle\psi_t\left|h_n\right| \psi_t\right\rangle$ for all Pauli strings in the problem Hamiltonian change along the evolution process as driven by pure non-Hermitian algorithm, non-Hermitian-Variational algorithm and QAOA.  It is easy to find that, compared with QAOA, the non-Hermitian algorithm and non-Hermitian-Variational algorithm give a much more stable and smooth variable evolution process, making the reduced recording process much more effective by recording the correlation properties $M_{h_n}=\left\langle\psi_t\left|h_n\right| \psi_t\right\rangle$ according to a confidence interval rather than the largest absolute value such as QAOA~\cite{bravyi2020obstacles}.

Since the correlation functions of non-Hermitian algorithm behave more predictably with strong time correlation and show less random temporal fluctuations as in the QAOA case, we also expect that besides the reduced recording method, the recursive idea constructed based on this non-Hermitian algorithm could outperform the original recursive QAOA~\cite{bravyi2020obstacles} by recursively set some correlation functions in confidence region to be $\pm 1$ and remove a few qubits once a time. This idea can also help solve the problem of measurements shots explosion for larger size systems, owing to the ability that the recursive non-Hermitian algorithm recursively decreases the system size during the simulation process. 

\section{Simulation models} \label{sec:models}
Here we define and give some details about the models we used in simulation experiments.

\subsection{1D transverse field Ising model} \label{sec:TFIMmodels}
The 1D TFIM Hamiltonian with periodic boundary condition reads
\begin{equation}
H_{TFIM}=J \sum_{\langle i j\rangle} Z_{i} Z_{j}+h_{X} \sum X_{i},
\end{equation}
where $X$ and $Z$ are the Pauli matrices and $\left< i,j \right>$ denotes pairs of interacting neighboring qubits, and $J, h_X$ is the coupling strength and transverse field strength, respectively. For all TFIM models we used in this work, we set $J=1/\sqrt{2},\  h_X=1/\sqrt{2}$. As TFIM at critical point is generally believed the most difficult to simulate.

\subsection{3-SAT model} \label{sec:3satmodels}
The 3-SAT problem is a paradigmatic example of a non-deterministic polynomial (NP) problem  \cite{Hogg2003}. A 3-SAT problem is defined by a logical statement involving \(n\) boolean variables \(b_i\). The logical statement consists of \(m\) clauses \(C_i\) in conjunction: \(C_{1} \wedge C_{2} \wedge \cdots \wedge C_{m}\). Each clause is a disjunction of 3 literals, where a literal is a boolean variable \(b_i\) or its negation \(\neg b_i\). For instance, a clause may read \(\left(b_{j} \vee \neg b_{k} \vee b_{l}\right)\). The task is to first decide whether a given 3-SAT problem is satisfiable; if so, then assign appropriate binary values to satisfy the logical statement.  
We can map a 3-SAT problem to a Hamiltonian for a set of qubits. Under this mapping, each binary variable $b_i$ is represented as a qubit state. Thus, an \(n\)-variable 3-SAT problem is mapped into a Hilbert space of dimension \(N = 2^n\). Furthermore, each clause of the logical statement is translated to a projector, projecting on the bitstrings that not satisfying each given clause. Hence, a logical statement with $m$ clauses may be translated to the following Hamiltonian,
\begin{equation}
H_{3-SAT}=\sum_{\alpha=1}^{m}\left|b_{j}^{\alpha} b_{k}^{\alpha} b_{l}^{\alpha}\right\rangle\left\langle b_{j}^{\alpha} b_{k}^{\alpha} b_{l}^{\alpha}\right|.
\end{equation}
Note 3-SAT problem intrinsically corresponds to a classical many-body spin-glass Hamiltonian with long range interaction.
Since the computational complexity is defined in terms of the worst-case performance, hard instances of 3-SAT have been intensively studied in the past. Following Ref \cite{nidari2005}, we focus on a particular set of 3-SAT instances, each is characterized with a unique solution and a ratio of $m/n=3$  in this work. We note that this ratio of $3$ is different from the phase-transition point $m/n \approx 4.2$  \cite{Kirkpatrick1994,Monasson1999} that has been intensively explored in studies that characterize the degrees of satisfiability of random 3-SAT problems. The subtle distinction is that the phase-transition point characterizes the notion of ``hardness" (with respect to the $m/n$ ratio) by averaging over 3-SAT instances having variable number of solutions. However, when the focus is to identify the most difficult 3-SAT instances having unique solution, it has been ``empirically" found that these instances tend to have an $m/n$ ratio lower than the phase-transition point.

\begin{table}  
\centering
\caption{3-SAT 5 Qubits}  
\begin{tabular*}{12cm}{lllll}  
\hline  
 IIIIZ : -0.25 &  IIIZI : 0.125 &   IIIZZ : -0.125 &  IIZII : 0.375 &  IIZIZ : 0.125 \\  
  IIZZI : -0.125 &  IIZZZ : 0.25 &  IZIII : -0.125 &  IZIIZ : 0.125 &  IZIZI : 0.5 \\  
  IZIZZ : -0.375 &  IZZII : -0.125 &  IZZZI : 0.25 &  ZIIII : 0.25 &  ZIIIZ : -0.125 \\
  ZIIZI : -0.25 &  ZIZII : -0.125 &  ZIZIZ : 0.125 &  ZIZZI : -0.125 &  ZZIII : -0.25 \\
  ZZIIZ : 0.25 &  ZZIZI : -0.125 &  ZZZII : -0.125 \\
\hline  
\end{tabular*} 
\label{tab:3sat5}
\end{table} 

\begin{table}  
\centering
\caption{3-SAT 8 Qubits}  
\begin{tabular*}{14cm}{lllll}  
\hline  
 IIIIIIIZ : -0.14453125 &  IIIIIIZI : -0.28515625 &  IIIIIIZZ : 0.01953125 &  IIIIIZII : -0.62890625 \\  IIIIIZIZ : 0.00390625 &  IIIIIZZI : 0.00390625 &  IIIIIZZZ : 0.01171875 &  IIIIZIII : -1.24609375 \\  IIIIZIIZ : -0.01953125 &  IIIIZIZI : 0.02734375 &  IIIIZIZZ : 0.00390625 &  IIIIZZII : -0.05078125 \\  IIIIZZIZ : 0.00390625 &  IIIIZZZI : 0.00390625 &  IIIIZZZZ : 0.02734375 &  IIIZIIII : -2.34765625 \\  IIIZIIIZ : -0.01171875 &  IIIZIIZI : -0.01171875 &  IIIZIIZZ : 0.01171875 &  IIIZIZII : -0.04296875 \\  IIIZIZIZ : -0.00390625 &  IIIZIZZI : 0.01171875 &  IIIZIZZZ : 0.01953125 &  IIIZZIII : 0.05859375 \\  IIIZZIIZ : 0.00390625 &  IIIZZIZI : -0.02734375 &  IIIZZIZZ : 0.01171875 &  IIIZZZII : -0.04296875 \\  IIIZZZIZ : 0.01171875 &  IIIZZZZI : -0.00390625 &  IIIZZZZZ : -0.01171875 &  IIZIIIII : -5.01953125 \\  IIZIIIIZ : -0.01171875 &  IIZIIIZI : -0.04296875 &  IIZIIIZZ : -0.00390625 &  IIZIIZII : -0.07421875 \\  IIZIIZIZ : 0.04296875 &  IIZIIZZI : -0.00390625 &  IIZIIZZZ : -0.01171875 &  IIZIZIII : -0.17578125 \\  IIZIZIIZ : -0.02734375 &  IIZIZIZI : 0.00390625 &  IIZIZIZZ : -0.03515625 &  IIZIZZII : -0.04296875 \\  IIZIZZIZ : -0.00390625 &  IIZIZZZI : 0.01171875 &  IIZIZZZZ : 0.01953125 &  IIZZIIII : -0.29296875 \\  IIZZIIIZ : 0.02734375 &  IIZZIIZI : -0.01953125 &  IIZZIIZZ : -0.01171875 &  IIZZIZII : -0.01953125 \\  IIZZIZIZ : 0.00390625 &  IIZZIZZI : 0.00390625 &  IIZZIZZZ : 0.05859375 &  IIZZZIII : 0.06640625 \\  IIZZZIIZ : -0.00390625 &  IIZZZIZI : -0.01953125 &  IIZZZIZZ : 0.00390625 &  IIZZZZII : -0.03515625 \\  IIZZZZIZ : 0.00390625 &  IIZZZZZI : -0.02734375 &  IIZZZZZZ : 0.01171875 &  IZIIIIII : -10.27734375 \\  IZIIIIIZ : -0.03515625 &  IZIIIIZI : 0.01171875 &  IZIIIIZZ : 0.06640625 &  IZIIIZII : 0.02734375 \\  IZIIIZIZ : 0.00390625 &  IZIIIZZI : 0.00390625 &  IZIIIZZZ : 0.01171875 &  IZIIZIII : -0.04296875 \\  IZIIZIIZ : -0.00390625 &  IZIIZIZI : -0.01953125 &  IZIIZIZZ : -0.01171875 &  IZIIZZII : 0.07421875 \\  IZIIZZIZ : 0.00390625 &  IZIIZZZI : -0.05859375 &  IZIIZZZZ : -0.00390625 &  IZIZIIII : -0.20703125 \\  IZIZIIIZ : 0.00390625 &  IZIZIIZI : 0.00390625 &  IZIZIIZZ : 0.05859375 &  IZIZIZII : 0.08203125 \\  IZIZIZIZ : -0.00390625 &  IZIZIZZI : 0.01171875 &  IZIZIZZZ : -0.01171875 &  IZIZZIII : 0.13671875 \\  IZIZZIIZ : -0.01171875 &  IZIZZIZI : 0.01953125 &  IZIZZIZZ : -0.00390625 &  IZIZZZII : 0.05078125 \\  IZIZZZIZ : 0.01171875 &  IZIZZZZI : -0.00390625 &  IZIZZZZZ : -0.01171875 &  IZZIIIII : 0.05859375 \\  IZZIIIIZ : 0.00390625 &  IZZIIIZI : 0.03515625 &  IZZIIIZZ : -0.01953125 &  IZZIIZII : 0.05078125 \\  IZZIIZIZ : -0.01953125 &  IZZIIZZI : -0.00390625 &  IZZIIZZZ : -0.04296875 &  IZZIZIII : 0.02734375 \\  IZZIZIIZ : 0.01953125 &  IZZIZIZI : -0.01171875 &  IZZIZIZZ : 0.01171875 &  IZZIZZII : -0.01171875 \\  IZZIZZIZ : -0.00390625 &  IZZIZZZI : 0.01171875 &  IZZIZZZZ : 0.01953125 &  IZZZIIII : -0.08984375 \\  IZZZIIIZ : 0.01171875 &  IZZZIIZI : 0.02734375 &  IZZZIIZZ : -0.02734375 &  IZZZIZII : 0.01171875 \\  IZZZIZIZ : 0.00390625 &  IZZZIZZI : 0.00390625 &  IZZZIZZZ : -0.00390625 &  IZZZZIII : 0.01953125 \\  IZZZZIIZ : 0.01171875 &  IZZZZIZI : -0.00390625 &  IZZZZIZZ : -0.01171875 &  IZZZZZII : 0.08984375 \\  IZZZZZIZ : 0.00390625 &  IZZZZZZI : -0.02734375 &  IZZZZZZZ : -0.01953125 &  ZIIIIIII : -21.35546875 \\  ZIIIIIIZ : -0.01953125 &  ZIIIIIZI : -0.01953125 &  ZIIIIIZZ : 0.00390625 &  ZIIIIZII : -0.09765625 \\  ZIIIIZIZ : -0.02734375 &  ZIIIIZZI : 0.05078125 &  ZIIIIZZZ : 0.02734375 &  ZIIIZIII : -0.16796875 \\  ZIIIZIIZ : -0.00390625 &  ZIIIZIZI : -0.00390625 &  ZIIIZIZZ : 0.00390625 &  ZIIIZZII : -0.00390625 \\  ZIIIZZIZ : -0.01171875 &  ZIIIZZZI : 0.00390625 &  ZIIIZZZZ : -0.00390625 &  ZIIZIIII : -0.20703125 \\  ZIIZIIIZ : 0.00390625 &  ZIIZIIZI : -0.01171875 &  ZIIZIIZZ : -0.01953125 &  ZIIZIZII : -0.05859375 \\  ZIIZIZIZ : -0.01953125 &  ZIIZIZZI : -0.01953125 &  ZIIZIZZZ : 0.01953125 &  ZIIZZIII : -0.09765625 \\  ZIIZZIIZ : -0.02734375 &  ZIIZZIZI : -0.01171875 &  ZIIZZIZZ : -0.00390625 &  ZIIZZZII : -0.10546875 \\  ZIIZZZIZ : 0.01171875 &  ZIIZZZZI : 0.04296875 &  ZIIZZZZZ : 0.00390625 &  ZIZIIIII : -0.72265625 \\  ZIZIIIIZ : 0.00390625 &  ZIZIIIZI : 0.01953125 &  ZIZIIIZZ : -0.00390625 &  ZIZIIZII : -0.05859375 \\  ZIZIIZIZ : 0.02734375 &  ZIZIIZZI : 0.02734375 &  ZIZIIZZZ : 0.01953125 &  ZIZIZIII : -0.05078125 \\  ZIZIZIIZ : 0.00390625 &  ZIZIZIZI : 0.01953125 &  ZIZIZIZZ : -0.01953125 &  ZIZIZZII : -0.01171875 \\  ZIZIZZIZ : -0.00390625 &  ZIZIZZZI : -0.00390625 &  ZIZIZZZZ : 0.00390625 &  ZIZZIIII : -0.01171875 \\  ZIZZIIIZ : -0.03515625 &  ZIZZIIZI : -0.00390625 &  ZIZZIIZZ : 0.00390625 &  ZIZZIZII : -0.01953125 \\  ZIZZIZIZ : -0.02734375 &  ZIZZIZZI : -0.01171875 &  ZIZZIZZZ : 0.04296875 &  ZIZZZIII : -0.07421875 \\  ZIZZZIIZ : 0.01171875 &  ZIZZZIZI : 0.01171875 &  ZIZZZIZZ : 0.03515625 &  ZIZZZZII : -0.08203125 \\  ZIZZZZIZ : -0.01171875 &  ZIZZZZZI : 0.03515625 &  ZIZZZZZZ : 0.01171875 &  ZZIIIIII : -1.73046875 \\  ZZIIIIIZ : 0.04296875 &  ZZIIIIZI : 0.07421875 &  ZZIIIIZZ : 0.00390625 &  ZZIIIZII : 0.10546875 \\  ZZIIIZIZ : -0.01171875 &  ZZIIIZZI : -0.02734375 &  ZZIIIZZZ : 0.04296875 &  ZZIIZIII : 0.26953125 \\  ZZIIZIIZ : 0.02734375 &  ZZIIZIZI : 0.05859375 &  ZZIIZIZZ : 0.00390625 &  ZZIIZZII : 0.04296875 \\  ZZIIZZIZ : 0.00390625 &  ZZIIZZZI : -0.01171875 &  ZZIIZZZZ : -0.01953125 &  ZZIZIIII : 0.48046875 \\  ZZIZIIIZ : -0.02734375 &  ZZIZIIZI : -0.07421875 &  ZZIZIIZZ : -0.01953125 &  ZZIZIZII : -0.01171875 \\  ZZIZIZIZ : -0.00390625 &  ZZIZIZZI : 0.02734375 &  ZZIZIZZZ : 0.00390625 &  ZZIZZIII : -0.16015625 \\  ZZIZZIIZ : -0.02734375 &  ZZIZZIZI : 0.01953125 &  ZZIZZIZZ : -0.00390625 &  ZZIZZZII : 0.03515625 \\  ZZIZZZIZ : 0.02734375 &  ZZIZZZZI : -0.03515625 &  ZZIZZZZZ : 0.01953125 &  ZZZIIIII : 1.27734375 \\  ZZZIIIIZ : 0.03515625 &  ZZZIIIZI : 0.01953125 &  ZZZIIIZZ : -0.00390625 &  ZZZIIZII : 0.05078125 \\  ZZZIIZIZ : -0.01953125 &  ZZZIIZZI : 0.01171875 &  ZZZIIZZZ : 0.00390625 &  ZZZIZIII : 0.13671875 \\  ZZZIZIIZ : 0.00390625 &  ZZZIZIZI : -0.01171875 &  ZZZIZIZZ : -0.01953125 &  ZZZIZZII : 0.00390625 \\  ZZZIZZIZ : 0.01171875 &  ZZZIZZZI : -0.01953125 &  ZZZIZZZZ : 0.01953125 &  ZZZZIIII : 0.11328125 \\  ZZZZIIIZ : -0.03515625 &  ZZZZIIZI : -0.03515625 &  ZZZZIIZZ : 0.00390625 &  ZZZZIZII : -0.00390625 \\  ZZZZIZIZ : -0.01171875 &  ZZZZIZZI : -0.02734375 &  ZZZZIZZZ : -0.00390625 &  ZZZZZIII : -0.13671875 \\  ZZZZZIIZ : -0.01953125 &  ZZZZZIZI : 0.01171875 &  ZZZZZIZZ : -0.02734375 &  ZZZZZZII : 0.02734375 \\     ZZZZZZIZ : 0.00390625 &  ZZZZZZZI : 0.01953125 &  ZZZZZZZZ : -0.00390625 \\
\hline  
\end{tabular*}  
\label{tab:3sat8}
\end{table} 
The detail information of  Hamiltonian we used in this work is presented below as in Table \ref{tab:3sat5} and Table \ref{tab:3sat8}.

\end{document}